\UseRawInputEncoding
\documentclass[5p]{elsarticle}

\usepackage[T1]{fontenc}
\usepackage[utf8]{inputenc}
\usepackage{amsmath}  
\usepackage{amssymb}
\usepackage{graphicx}
\usepackage{caption}
\usepackage{subcaption} 
\usepackage{appendix}

\usepackage{amsmath}
\usepackage{mathtools}
\usepackage[version=4]{mhchem}
\usepackage{enumitem}
\usepackage{upgreek}
\usepackage{array}
\usepackage{tabularx}
\usepackage{threeparttable}
\usepackage[
mode=text,  
range-phrase=\textendash,
range-units=single,
retain-explicit-plus,
]{siunitx}
\usepackage{tablefootnote}
\usepackage{multirow}
\usepackage{xcolor, color, soul}
\sethlcolor{yellow}
\usepackage{booktabs, caption, makecell}
\usepackage[colorlinks,citecolor=red,urlcolor=blue,bookmarks=false,hypertexnames=true]{hyperref} 
\hypersetup{colorlinks = true,linkcolor = blue,anchorcolor =red,citecolor = blue,filecolor = red,urlcolor = red,
           pdfauthor=author}
\newcolumntype{C}[1]{>{\centering\let\newline\\\arraybackslash\hspace{0pt}}m{#1}}
\DeclareUnicodeCharacter{2212}{\textendash}
\DeclareUnicodeCharacter{03B1}{$\alpha$}
\DeclareUnicodeCharacter{03B2}{$\beta$}
\DeclareUnicodeCharacter{00E1}{\'{a}}
\DeclareUnicodeCharacter{223C}{$\sim$}
\biboptions{sort&compress}
\definecolor{lightblack}{rgb}{0.8, 0.8, 0.8}
\definecolor{responsered}{RGB}{0,0,0}
\usepackage{orcidlink}

\begin{document}
\sloppy
\begin{frontmatter}


\title{Hydrogen diffusion in \texorpdfstring{TiCr\textsubscript{2}H\textsubscript{\textit{x}}}{TiCr2Hx} Laves phases: A combined \textit{ab initio} and machine-learning-potential study}

\affiliation[inst1]{
organization={Institute for Materials Science, University of Stuttgart},
addressline={Pfaffenwaldring 55},
postcode={70569},
city={Stuttgart},
country={Germany},
}
\affiliation[inst2]{
organization={Interdisciplinary Centre for Advanced Materials Simulation (ICAMS), Ruhr-Universität Bochum},
postcode={44801},
city={Bochum},
country={Germany},
}

\affiliation[inst2b]{
organization={Department of Computational Materials Design, Max-Planck-Institut for Sustainable Materials},
postcode={40237},
city={D\"usseldorf},
country={Germany},
}

\affiliation[inst3]{
organization={WPI, International Institute for Carbon-Neutral Energy Research (WPI-I2CNER), Kyushu University},
postcode={819-0395},
city={Fukuoka},
country={Japan},
}
\affiliation[inst4]{
organization={Department of Automotive Science, Graduate School of Integrated Frontier Sciences, Kyushu University},
postcode={819-0395},
city={Fukuoka},
country={Japan},
}
\affiliation[inst5]{
organization={Mitsui Chemicals, Inc. -.Carbon-Neutral Research Center (MCI-CNRC), Kyushu University},
postcode={819-0395},
city={Fukuoka},
country={Japan},
}

\author[inst1]{Pranav Kumar\corref{cor1}\,\orcidlink{0000-0002-3661-5870}}
\ead{pranav.kumar@imw.uni-stuttgart.de}
\author[inst2,inst2b]{Fritz Körmann\,\orcidlink{0000-0003-3050-6291}}
\author[inst3,inst4,inst5]{Kaveh Edalati\,\orcidlink{0000-0003-3885-2121}}
\author[inst1]{Blazej Grabowski\,\orcidlink{0000-0003-4281-5665}}
\author[inst1]{Yuji Ikeda\corref{cor1}\,\orcidlink{0000-0001-9176-3270}}
\ead{yuji.ikeda@imw.uni-stuttgart.de}
\cortext[cor1]
{Corresponding authors}             
            


\begin{abstract}
The kinetics of hydrogen diffusion in C15 cubic and C14 hexagonal \ce{TiCr2H_x} (0~$<$~\textit{x}~$\le$~4) Laves-phase hydrogen storage alloys is investigated with density functional theory (DFT) and machine learning interatomic potentials (MLIPs).
Generalized solid-state nudged elastic band calculations are conducted based on DFT for all symmetrically inequivalent paths between the first-nearest-neighbor face-sharing interstitial sites.
The hydrogen migration barriers are substantially higher for the paths that require breaking a Ti--H bond than for those that require breaking a Cr--H bond.
Molecular dynamics (MD) simulations with the MLIPs also demonstrate that hydrogen migration occurs more frequently within the hexagonal rings made of the \ce{A2B2} interstitial paths, each requiring the breaking of Cr--H bonds, than along the inter-ring paths.
The diffusion coefficients of hydrogen obtained from the MD simulations reveal a non-monotonic dependence on hydrogen concentration, which is more pronounced at lower temperatures.
Time-averaged radial distribution functions of hydrogen further show that hydrogen avoids face-sharing positions during diffusion and that the hydrogen occupancy at the second-nearest-neighbor edge-sharing positions increases with increasing hydrogen concentration.
The diffusion coefficients of hydrogen within \qtyrange{400}{1000}{K} follow an Arrhenius relationship, with activation barriers consistent with most experimental values.
One-order of magnitude overestimation of diffusion coefficients compared with some experiments suggests a substantial impact of hydrogen trapping by defects such as Cr vacancies and Ti anti-sites in non-stoichiometric \ce{TiCr2} in experiments.
\end{abstract}

\begin{graphicalabstract}
\includegraphics[width=\textwidth]{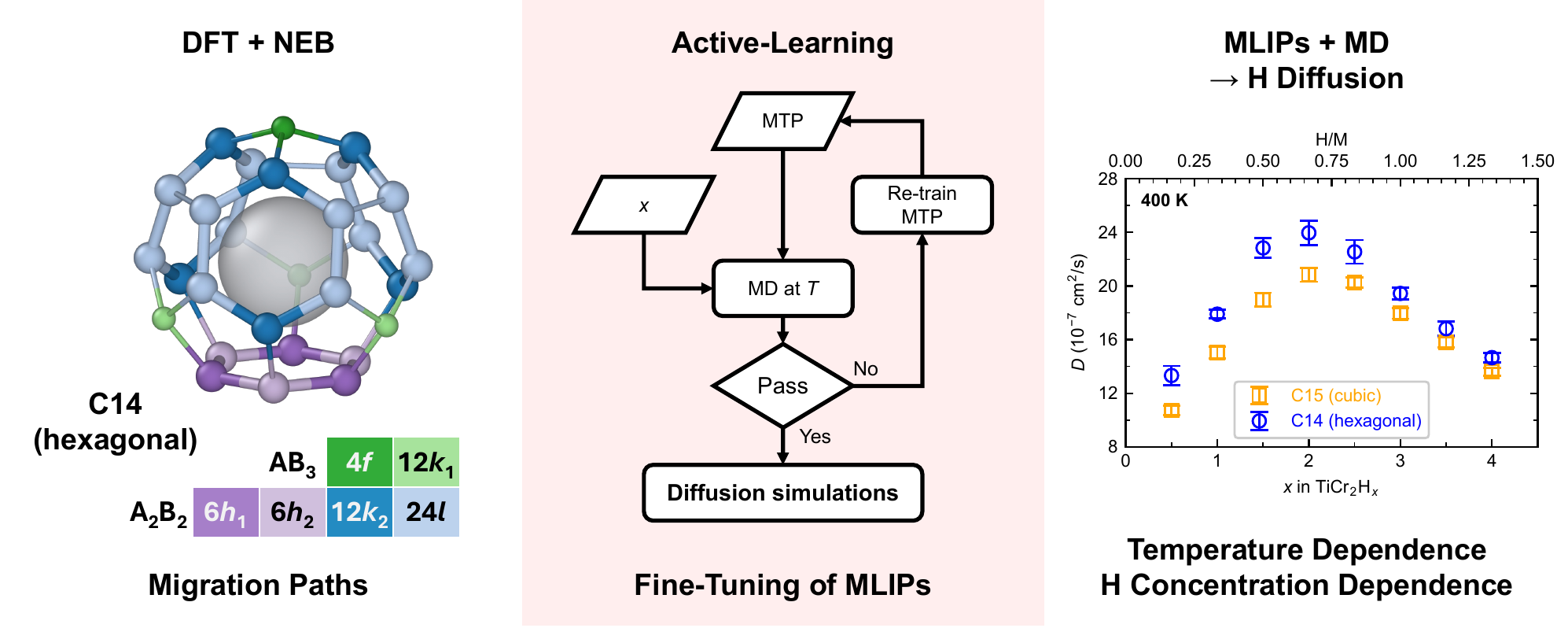}
\end{graphicalabstract}






\end{frontmatter}


\section{Introduction}
\label{sec:Introduction}

\ce{AB2} Laves-phase alloys have emerged as promising materials for solid-state hydrogen storage~\cite{stein_laves_2021,yartys_laves_2022}. Various experimental studies have investigated the hydrogen absorption kinetics of Laves-phase alloys to evaluate their hydrogen storage performance~\cite{edalati_reversible_2020,mohammadi_high-entropy_2022,ding_interface_2023,strozi_tuning_2023,dangwal_machine_2024,andrade_microstructural_2025}. Among these alloys, \ce{Ti}-based systems have been extensively studied for their hydrogen absorption and storage capabilities \cite{osumi_hydrogen_1983,manickam_optimization_2015,villeroy_influence_2006}. In particular, the hydrogenation characteristics of \ce{TiCr2} has been thoroughly investigated~\cite{johnson_reaction_1978,johnson_reaction_1980,mcgrath_stoichiometry_2024}, providing valuable insights into alloying strategies designed to enhance hydrogenation performance~\cite{agresti_reaction_2009,jiang_influence_2022}.

The diffusion of hydrogen plays a critical role in determining the kinetic performance of hydrogen-storage materials and has therefore been the subject of extensive investigations.
Specifically for \ce{TiCr2} in its Laves phases, proton relaxation measurements with nuclear magnetic resonance (NMR) by Bowman~\textit{et al.}~\cite{bowman_diffusion_1983} revealed a wide distribution of activation barriers.
The C15 cubic structure at compositions of \ce{TiCr_{1.8}H_{x}} with $x = \numlist{0.55;2.58}$ exhibits activation barriers for hydrogen diffusion of approximately \qtylist{0.2;0.3}{eV}, respectively, near room temperature. The C14 hexagonal structure at similar compositions of \ce{TiCr_{1.9}H_{x}} with $x = \numlist{0.63;2.85}$ shows activation barriers of approximately \qtylist{0.2;0.4}{eV}, respectively. These results imply a dependence of the activation barriers on hydrogen concentration.
Quasi-elastic neutron scattering measurements by Campbell \textit{et al.}~\cite{campbell_quasi-elastic_1999} revealed an activation barrier for hydrogen diffusion of \qty{0.24}{eV} in C15 \ce{TiCr_{1.8}H_{0.43}} at \qtyrange{313}{442}{K},
comparable to the value obtained from the NMR study for the same C15 phase with a similar hydrogen concentration~\cite{bowman_diffusion_1983}.
Mazzolai \textit{et al.}~\cite{mazzolai_hydrogen_2009} investigated hydrogen diffusion in \ce{TiCr_{1.78}H_{x}} for a dilute hydrogen concentration ($x \approx 0.017$) using internal friction and hydrogen absorption measurements.
They reported that the activation barrier can reach up to \qty{0.60}{eV} at \qtyrange{660}{1200}{K}, i.e., a value substantially larger than those obtained by the other experiments~\cite{hiebl_proton_1982,bowman_diffusion_1983,campbell_quasi-elastic_1999}.
It is challenging to interpret the scatter in the activation barriers from experiments alone.

Atomistic simulations based on density functional theory (DFT) can provide further insight into the hydrogen diffusion properties.
Indeed, several computational studies have quantified the energetics and migration barriers of hydrogen in the \ce{TiCr2} Laves phases~\cite{li_hydrogen_2009, miwa_path_2019, jiang_influence_2022, loh_substitutional_2023}.
However, most existing computational works have focused on dilute solution energies and provided only a limited analysis of migration barriers between \ce{A2B2} sites.
Moreover, quantitative assessments of hydrogen diffusion coefficients in \ce{TiCr2} Laves phases across a wide range of hydrogen concentrations are largely absent from prior computational studies.
This is primarily due to the high computational cost of DFT.
Accurately resolving hydrogen distribution and interactions requires large length and time scales, which makes \textit{ab initio} molecular dynamics (MD) simulations computationally prohibitive.

Recent advancements in machine learning interatomic potentials (MLIPs) now allow MD simulations at larger spatial and temporal scales while retaining near-DFT accuracy.
MD simulations employing MLIPs have indeed been successfully applied to the investigation of atomic diffusion, such as hydrogen~\cite{wang_molecular_2020,kwon_accurate_2023,qi_robust_2024,yu_hydrogen_2024,angeletti_hydrogen_2025,morrison_long_2025} and lithium~\cite{ou_atomistic_2024,ou_non-arrhenius_2025}.
In our previous study~\cite{KUMAR2025121319}, we developed MLIPs for studying hydrogenated \ce{TiCr2} Laves phases. The developed MLIPs accurately account for a wide range of hydrogen concentrations while capturing the precise energetics involved. 

In the present study, we systematically investigate hydrogen migration and diffusion in C15 and C14 \ce{TiCr2} using DFT and MLIPs. With DFT, we quantify hydrogen migration barriers for all possible symmetrically inequivalent paths between face-sharing interstitial sites.
MD simulations are performed to investigate diffusion coefficients of hydrogen using MLIPs trained on DFT data.
In particular, we analyze how the hydrogen diffusion coefficients depend on hydrogen concentration and temperature in both the C15 and the C14 phases.

\section{Methodology}
\label{sec:methodology}

\subsection{Crystal structures and interstitial sites}
\label{sec:methodology:models}

\begin{table}[!tb]
\centering
\scriptsize
\caption{Fractional coordinates of the metal atoms (A and B) and the interstitial sites (B\textsubscript{4}, AB\textsubscript{3}, and A\textsubscript{2}B\textsubscript{2}) in the \ce{AB2} Laves phases.
The C15 cubic and the C14 hexagonal Laves phases show symmetries of the space groups $Fd\bar3m$ (227) and $P6_3/mmc$ (194), respectively.
The positions in the C15 phase are given with the origin choice 2, i.e., the origin set at the inversion center.
The subscripts at the Wyckoff letters distinguish the interstitial sites and follow the notation of Shoemaker and Shoemaker~\cite{shoemaker_concerning_1979}.
The values in bold are constrained by symmetry, while the others refer to the ideal case of a close-packed hard-sphere model with an atomic-radius ratio of $r_\mathrm{A}/r_\mathrm{B} = (3/2)^{1/2} \approx 1.225$ \cite{thoma_geometric_1995} and a $c/a$ ratio of $(8/3)^{1/2} \approx 1.633$ for the C14 phase.
The positions of the interstitial sites are given as the geometric centers of the surrounding four metal atoms.}
\label{tab:positions_ideal}
\begin{tabular}{ccc*3{C{12mm}}}
\toprule
    & Type & Wyckoff & $x$ & $y$ & $z$ \\
\midrule
C15
& \ce{A}    & $8a$  & $\mathbf{1/8}$ & $\mathbf{1/8}$ & $\mathbf{1/8}$ \\
& \ce{B}    & $16d$ & $\mathbf{1/2}$ & $\mathbf{1/2}$ & $\mathbf{1/2}$ \\
\cmidrule{2-6}
& \ce{B4}   & $8b$  & $\mathbf{3/8}$ & $\mathbf{3/8}$ & $\mathbf{3/8}$ \\
\cmidrule{2-6}
& \ce{AB3}  & $32e$ & $9/32$ & $9/32$ & $9/32$ \\
\cmidrule{2-6}
& \ce{A2B2} & $96g$ & $5/16$ & $5/16$ & $1/8$ \\
\midrule
C14 & \ce{A}    &  $4f$  & $\mathbf{1/3}$ & $\mathbf{2/3}$ & $1/16$ \\
    & \ce{B}    &  $2a$  & $\mathbf{0}$ & $\mathbf{0}$ & $\mathbf{0}$ \\
    &           &  $6h$  & $-1/6$ & $-2/6$ & $\mathbf{1/4}$ \\
\cmidrule{2-6}
    & \ce{B4}   & $4e$   & $\mathbf{0}$ & $\mathbf{0}$ & $3/16$ \\
\cmidrule{2-6}
    & \ce{AB3}  & $4f$   & $\mathbf{1/3}$ & $\mathbf{2/3}$ & $43/64$ \\
    &           & $12k_1$& $1/8$ & $1/4$ & $23/64$ \\
\cmidrule{2-6}
    & \ce{A2B2}
      & $6h_1$  & $ 5/24$ & $ 5/12$ & $\mathbf{1/4}$ \\
    & & $6h_2$  & $11/24$ & $11/12$ & $\mathbf{1/4}$ \\
    & & $12k_2$ & $13/24$ & $13/12$ & $1/8$ \\
    & & $24l$   & $ 1/24$ & $ 1/ 3$ & $9/16$ \\
\bottomrule
\end{tabular}
\end{table}

\begin{figure}[!tb]
\centering
\includegraphics{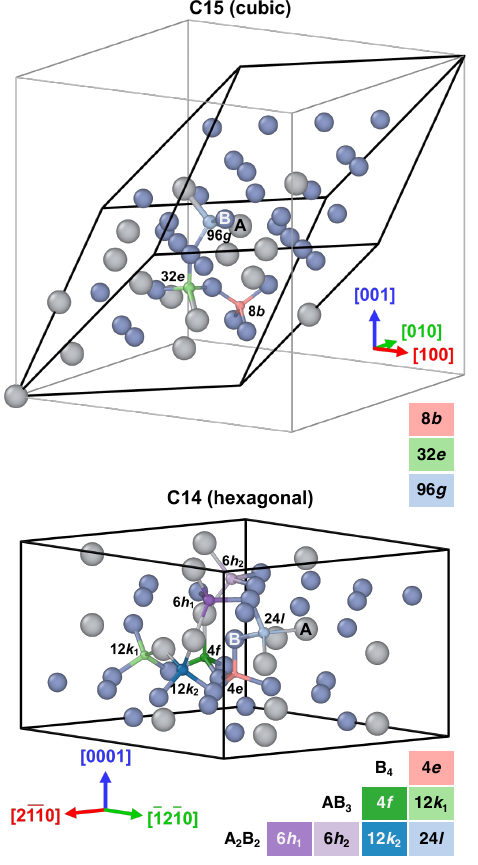}
\caption{48-metal-atom cells of the C15 cubic and the C14 hexagonal Laves phases together with symmetrically distinct interstitial sites.
Ti--H bonds are depicted thicker than Cr--H bonds to indicate that the former requires higher energies to break during hydrogen migration (cf.~Sec.~\ref{sec:results:migration_energies}).
The gray-colored cell for C15 represents a cubic supercell surrounding the 48-metal-atom cell.}
\label{fig:wyckoff}
\end{figure}

The \ce{Ti-Cr} system reveals the C15 and the C14 Laves phases near the \ce{TiCr2} composition in experiments~\cite{chen_stoichiometry_1994,murray_crti_1981,sluiter_phase_1991,mebed_computer_1998}.
The low-temperature C15 phase has a cubic symmetry belonging to the space group 227 (\textit{Fd}\={3}\textit{m}). The high-temperature C14 phase has a hexagonal symmetry belonging to the space group 194 (\textit{P}6\textsubscript{3}/\textit{mmc}).
Table~\ref{tab:positions_ideal} summarizes the ideal positions of the metal atoms as well as the interstitial sites in the C15 and the C14 Laves phases.
In such \ce{AB2} Laves phases, hydrogen atoms may be accommodated in three chemically distinct tetrahedral interstitial site types, i.e., B$_4$, AB$_3$, and A$_2$B$_2$, characterized by four surrounding metal atoms. The numbers of symmetrically in-equivalent interstitial site sublattices in the C15 cubic and C14 hexagonal Laves phases are three and seven, respectively.

\subsection{Hydrogen migration barriers from DFT}
\label{sec:methodology:NEB}

DFT calculations were conducted using VASP\,6.2.1~\cite{furthmuller_dimer_1996,kresse_efficiency_1996,kresse_ultrasoft_1999}
with a plane-wave basis and with the projector augmented-wave (PAW) method~\cite{blochl_projector_1994}.
The exchange--correlation energy was obtained using the generalized gradient approximation (GGA) in the Perdew--Burke--Ernzerhof (PBE) form~\cite{perdew_generalized_1996}.
The valence electron configurations in the PAW potentials were $[\text{Ne}] 3s^2 3p^6 4s^2 3d^2$, $[\text{Ar}] 4s^1 3d^5$ and $1s^1$ for Ti, Cr, and H atoms, respectively.
The plane-wave energy cutoff was set to 400\,eV.
Unless otherwise specified, all calculations were conducted with supercell models of a $2 \times 2 \times 2$ expansion of the primitive C15 unit cell and a $2 \times 2 \times 1$ expansion of the C14 unit cell, each containing 48 metal atoms, as visualized in Figure~\ref{fig:wyckoff}.
\textcolor{responsered}{(The impact of the supercell size on the binding and migration energies of hydrogen was examined for various expansions to ensure convergence of the DFT values; see Sec.~\textcolor{violet}{S1} in the supplementary materials (SM) for details.)}
The reciprocal space was sampled with a $\Gamma$-centered $k$-point mesh of $4 \times 4 \times 4$ for both the C15 and the C14 supercell models along with the Methfessel--Paxton smearing~\cite{methfessel_high-precision_1989} with a width of 0.1\,eV.
Electronic self-consistent-field iterations were performed until the energy convergence threshold of $1 \times 10^{-5}$\,eV was reached.
Spin-polarization was not considered, as it does not affect the energies and the forces on atoms at any hydrogen concentration relevant to the present study~\cite{KUMAR2025121319}.

In the dilute hydrogen limit (single hydrogen in the simulation cell), there are a total of 4 and 11 symmetrically distinct migration paths between first-nearest-neighbor face-sharing tetrahedral interstitial sites for the C15 cubic and C14 hexagonal phases, respectively, as detailed in Sec.~\ref{sec:results:migration_energies}. The minimum-energy paths for these migration paths were calculated using the generalized solid-state nudged elastic band (G-SSNEB) method~\cite{sheppard_generalized_2012}, as implemented in the VTST software~\cite{henkelman_climbing_2000,henkelman_improved_2000}.
For the initial and final configurations of each minimum-energy path, a single hydrogen atom was placed at the corresponding interstitial site in the 48-metal-atom supercell, and then all atomic positions and cell vectors were optimized to their ground state, resulting in volumetric distortion relative to pristine \ce{TiCr2}.
Between the initial and final states, 11 linearly interpolated intermediate images were generated and connected by a spring constant of 5 eV/\AA\textsuperscript{2}. The atomic positions and cell shapes of the intermediate images were further optimized using the quick-min method \cite{sheppard_optimization_2008} with a time step of \qty{0.1}{fs} and a force convergence criterion of \textcolor{responsered}{\qty{5e-3}{eV/\AA}}.

\subsection{Fine-tuning of MLIPs for diffusion}
\label{sec:methodology:MLIPs}

In the present study, we started with two previously developed MLIPs for the \ce{TiCr2} Laves phases with hydrogen~\cite{KUMAR2025121319}, corresponding to the C14 phase (C14-MTP) and the C15 phase (C15-MTP).
These MLIPs, specifically moment tensor potentials (MTPs)~\cite{shapeev_moment_2016}, reproduce the DFT energies with an accuracy of approximately \qty{3}{meV/atom} for a wide range of hydrogen concentrations, i.e., \ce{TiCr2H_x} where $0 \le x \le 6$.

The original C14-MTP, however, required refinement in order to achieve stable and accurate diffusion calculations at high hydrogen concentrations of $x \geq 3.5$.
Specifically, C14-MTP occasionally generated unrealistically large forces, which caused hydrogen atoms to undergo unphysical displacements.
To address this issue, we implemented the following iterative fine-tuning protocol based on active learning. 
\textcolor{responsered}{We started from a configuration with a hydrogen concentration of $x = 3.5$, with hydrogen atoms initially at the geometric centers of the interstitial sites. These configurations were then relaxed---allowing both atomic positions and the cell shape to adjust---using the existing C14-MTP potential. The relaxed structures were subsequently heated to \qty{1000}{K} under the \textit{NPT} ensemble. Once the volume and cell shape reached thermal equilibrium, the simulations were switched to the \textit{NVE} ensemble.
During the \textit{NVE} MD simulation, configurations with the extrapolation grades~\cite{podryabinkin_active_2017,gubaev_accelerating_2019,podryabinkin_mlip-3_2023} above 2.1 were continuously extracted and saved.}
The simulation was terminated once the extrapolation grade exceeded 10.
The saved configurations were re-filtered to identify additional configurations required to construct the active set~\cite{podryabinkin_active_2017,gubaev_accelerating_2019,podryabinkin_mlip-3_2023} based on D-optimality.
The energies, forces, and stresses of these additional configurations were computed with single-point DFT calculations, and the thus labeled configurations were incorporated into the original training dataset.
The MTP was then retrained on this augmented dataset.
\textcolor{responsered}{This active-learning process was repeated with the updated MTP until no additional configurations exhibited extrapolation grades above a selection threshold of 2.1 during an MD simulation of \qty{0.1}{ns} for hydrogen concentrations $x=3.5$ and $x=4.0$.}
Through this fine-tuning, 101 new configurations were added to the training dataset, increasing its size from \num{18902} to \num{19003} structures.
\textcolor{responsered}{The fine-tuned C14-MTP achieves RMSEs of \qty{2.87}{meV/atom} for energies and 0.095~eV/\AA\ for forces on the training set, comparable to the original C14-MTP.}
At the end, the fine-tuned C14-MTP successfully resolved the stability issues of the original potential, enabling robust and reliable MD simulations of hydrogen diffusion at high hydrogen concentrations.

In contrast, the previously developed C15-MTP~\cite{KUMAR2025121319} for modeling the C15 phase exhibited stable MD simulations up to \textit{x} = 4 in TiCr\textsubscript{2}H\textsubscript{\textit{x}}, and therefore no fine-tuning was required.

\subsection{Diffusion coefficients of hydrogen from MD}
\label{sec:methodology:diffusion}

The hydrogen diffusion coefficient was computed from the mean squared displacement (MSD) of hydrogen atoms in an MD simulation.
The MSD was evaluated as  
\begin{equation}
    \text{MSD}(t) = \frac{1}{N_\mathrm{H}} \sum_{i=1}^{N_\mathrm{H}} \left| \mathbf{r}_i(t) - \mathbf{r}_i(0) \right|^2,
\end{equation}
where $\mathbf{r}_i(t)$ denotes the position of atom \textit{i} at time $t$, and
$N_\mathrm{H}$ is the number of hydrogen atoms.
The MSD can be assumed to depend on \textit{t} as
\begin{equation}
    \label{eq:msd_alpha}
    \text{MSD}(t) \propto t^\alpha.
\end{equation}
When the MD simulation time is sufficiently long, the system enters the normal-diffusion regime, and $\alpha$ becomes approximately equal to 1, i.e., the MSD depends linearly on \textit{t}.
The diffusion coefficient $D$ is then obtained from the long-time limit of the MSD according to the Einstein equation~\cite{einstein_uber_1905} as  
\begin{equation}
    \label{eq:D_alpha}
    D = \lim_{t \to \infty} \frac{\text{MSD}(t)}{6t}.
\end{equation}

The MD simulations were conducted in simulation cells with 6000 metal atoms obtained by a $5 \times 5 \times 5$ expansion of the original 48-metal-atom supercell models, with the size of the simulation cells determined based on a convergence test for the diffusion coefficient (\ref{sec:D_vs_supercell}).
In the MD simulations, hydrogen atoms were initially distributed among the interstitial sites randomly. Each system was subsequently equilibrated using the isothermal-isobaric (\textit{NPT}) ensemble with a time step of \qty{1}{fs} and a duration of \qtyrange{50}{1500}{ps} depending on temperature.
The simulations were then switched to the microcanonical (\textit{NVE}) ensemble for over \qty{1}{ns} up to \qty{5}{ns}. Temperatures of \qtylist{250;400;500;700;1000}{K} and eight hydrogen concentrations in $0<x\leq4$ for \ce{TiCr2H_x} were considered.
To ensure converged statistics, the MD simulations were conducted twice for \qty{250}{K} and five times for the other temperatures. The $\alpha$ values in Eq.~\eqref{eq:msd_alpha} were first examined to see whether the MD simulation time reached the normal-diffusion regime.
As detailed in Sec.~\ref{sec:results:MSD}, the MD simulations reached the normal-diffusion regime at all the temperatures except for \qty{250}{K}, and therefore the diffusion coefficients were obtained at temperatures of \qty{400}{K} and above following Eq.~\eqref{eq:D_alpha}.

The MD simulations were carried out using LAMMPS \cite{thompson_lammps_2022} with the MTPs.
In most simulations, the MLIP-2 interface~\cite{novikov_mlip_2020} was used.
For simulations at \qty{400}{K}, another recently developed implementation \cite{meng_kokkos-accelerated_2025} was employed, yielding approximately a twofold speedup over the MLIP-2 interface on CPUs.

The C14 Laves phase has hexagonal symmetry, and therefore its hydrogen diffusion coefficients are expected to be anisotropic. As detailed in \ref{sec:anisotropy}, however, the anisotropy in C14 \ce{TiCr2H_x} is found to be negligible at temperatures of \qty{400}{K} and above, and therefore averaged isotropic diffusion coefficients are discussed in the main text.

In general, quantum effects substantially affect hydrogen diffusion in various systems.
In most cases, quantum tunneling reduces the activation barriers at low temperatures, which causes a temperature-induced transition of the dominant diffusion scheme, resulting in a non-Arrhenius trend~\cite{lauhon_direct_2000,zheng_quantum_2006,kimizuka_effect_2011,di_stefano_first-principles_2015}.
Specifically for \ce{TiCr2} in the Laves phases, Miwa and Asahi~\cite{miwa_path_2019} investigated hydrogen diffusion in C15 \ce{TiCr2H_{1.3}} at room temperature using the path integral hybrid Monte Carlo method \cite{duane_hybrid_1987,tuckerman_efficient_1993,mehlig_hybrid_1992} in combination with an MLIP trained with DFT data and elucidated the lowering of the activation barriers compared with classical MD results.
On the other hand, Mazzolai~\textit{et al.}~\cite{mazzolai_hydrogen_2009} reported that, in experiments, quantum effects on hydrogen diffusion in C14 \ce{TiCr_{1.78}H_{x}} at a dilute hydrogen concentration ($x \approx 0.017$) are prominent around \qty{120}{K} while moderate around room temperature.
As our primary focus in the present work is on how the diffusion coefficients depend on hydrogen concentration, we performed classical MD simulations mainly at temperatures above room temperature, where quantum effects play only a minor role.

\section{Results and Discussion}
\label{sec:Results}

\subsection{Interstitial site preferences in the \ce{TiCr2} Laves phases}
\label{sec:results:solution_energies}

Table~\ref{tab:binding_energies} summarizes the binding energies of single isolated hydrogen atoms in the C15 cubic and the C14 hexagonal Laves phases reproduced from our previous DFT study~\cite{KUMAR2025121319}.
In both the C15 and the C14 phases, hydrogen solubility is highest at the \ce{A2B2}-type interstitial sites, followed by the \ce{AB3}- and \ce{B4}-type sites in decreasing order. This preference is consistent with previous computational and experimental studies, which report that hydrogen predominantly occupies the \ce{A2B2} sites in \ce{TiCr2}~\cite{li_hydrogen_2009,li_achieving_2019} and other Laves-phase alloys~\cite{gesari_hydrogen_2010,merlino_dft_2016,jiang_influence_2022,yartys_laves_2022}.
In the C14 hexagonal phase, the tetrahedral interstices are further classified into seven symmetrically distinct ones.
Among the four \ce{A2B2} sites in C14, the 6\textit{h}\textsubscript{2} site is the most favorable for hydrogen, followed by the other three sites.
\textcolor{responsered}{The site preferences predicted by the present MTPs are consistent with the respective DFT results (see Sec.~\textcolor{violet}{S3} in the SM).}

\begin{table}[tbp]
\centering
\scriptsize
\caption{Binding energies \textit{E}\textsubscript{b} of a single hydrogen atom at the interstices of the C15 cubic and the C14 hexagonal \ce{TiCr2} Laves phases obtained with DFT~\cite{KUMAR2025121319}.}
\label{tab:solution_energy_single}
\begin{tabular}{
ccc
S[table-format=5.1,retain-explicit-plus]
}
\toprule
& Type & Site  & \multicolumn{1}{c}{\textit{E}\textsubscript{b} (meV/H)} \\
\midrule
C15                  & \ce{B4}   & $8b$     & +599 \\
\cmidrule{2-4}
                     & \ce{AB3}  & $32e$    &  -16 \\
\cmidrule{2-4}
                     & \ce{A2B2} & $96g$    & -217 \\
\midrule
C14                  & \ce{B4}   & $4e$     & +566 \\
\cmidrule{2-4}
                     & \ce{AB3}  & $4f$     & +077 \\
                     &           & $12k_1$  &  -19 \\
\cmidrule{2-4}
                     & \ce{A2B2} & $6h_1$   & -170 \\
                     &           & $6h_2$   & -249 \\
                     &           & $12k_2$  & -173 \\
                     &           & $24l$    & -185 \\
\bottomrule
\end{tabular}
\label{tab:binding_energies}
\end{table}

\subsection{Hydrogen migration barriers}
\label{sec:results:migration_energies}

\begin{table*}[tb]
\centering
\scriptsize
\caption{Hydrogen migration energies \textit{E}\textsubscript{mig} of all symmetrically distinct paths in the \ce{TiCr2} Laves phases obtained using DFT.}
\label{tab:migration}
\begin{tabular}{c c@{ }c c@{ }c c c c *{2}{S[table-format=4.0,color=responsered]}}
\toprule
&
\multicolumn{2}{c}{\multirow{2}{*}{\textit{S}\textsubscript{1}}} &
\multicolumn{2}{c}{\multirow{2}{*}{\textit{S}\textsubscript{2}}} &
\multirow{2}{*}{Face} &
\multicolumn{2}{c}{Bond breaking} &
\multicolumn{2}{c}{\textit{E}\textsubscript{mig} (meV)} \\
\cmidrule(lr){7-8}
\cmidrule(lr){9-10}
&&&&&&
\multicolumn{1}{c}{$S_1 \to S_2$} &
\multicolumn{1}{c}{$S_2 \to S_1$} &
\multicolumn{1}{c}{$S_1 \to S_2$} &
\multicolumn{1}{c}{$S_2 \to S_1$} \\
\midrule
C15
 & $8b$  & (\ce{B4})     & $32e$   & (\ce{AB3})  & \ce{B3}  & B--H & A--H & 186 & 801 \\
 & $32e$ & (\ce{AB3})    & $96g$   & (\ce{A2B2}) & \ce{AB2} & B--H & A--H &  90 & 290 \\
 & $96g$ & (\ce{A2B2})   & $96g$   & (\ce{A2B2}) & \ce{AB2} & A--H & A--H & 285 & 285 \\
 & $96g$ & (\ce{A2B2})   & $96g$   & (\ce{A2B2}) & \ce{A2B} & B--H & B--H & 123 & 123 \\
\midrule
C14
 & $4e$    & (\ce{B4})   & $4e$    & (\ce{B4})   & \ce{B3}  & B--H & B--H & 154 & 154 \\
 & $4e$    & (\ce{B4})   & $12k_1$ & (\ce{AB3})  & \ce{B3}  & B--H & A--H & 154 & 739 \\
 & $4f$    & (\ce{AB3})  & $4f$    & (\ce{AB3})  & \ce{B3}  & A--H & A--H & 630 & 630 \\
 & $4f$    & (\ce{AB3})  & $12k_2$ & (\ce{A2B2}) & \ce{AB2} & B--H & A--H & 111 & 361 \\
 & $12k_1$ & (\ce{AB3})  & $6h_1$  & (\ce{A2B2}) & \ce{AB2} & B--H & A--H &  98 & 249 \\
 & $12k_1$ & (\ce{AB3})  & $24l$   & (\ce{A2B2}) & \ce{AB2} & B--H & A--H &  97 & 263 \\
 & $6h_1$  & (\ce{A2B2}) & $6h_2$  & (\ce{A2B2}) & \ce{A2B} & B--H & B--H &  87 & 166 \\
 & $6h_2$  & (\ce{A2B2}) & $12k_2$ & (\ce{A2B2}) & \ce{AB2} & A--H & A--H & 280 & 203 \\
 & $12k_2$ & (\ce{A2B2}) & $24l$   & (\ce{A2B2}) & \ce{A2B} & B--H & B--H & 111 & 124 \\
 & $24l$   & (\ce{A2B2}) & $24l$   & (\ce{A2B2}) & \ce{AB2} & A--H & A--H & 313 & 313 \\
 & $24l$   & (\ce{A2B2}) & $24l$   & (\ce{A2B2}) & \ce{A2B} & B--H & B--H & 100 & 100 \\
\bottomrule
\end{tabular}

\end{table*}

\begin{figure}[tb]
\centering
\includegraphics{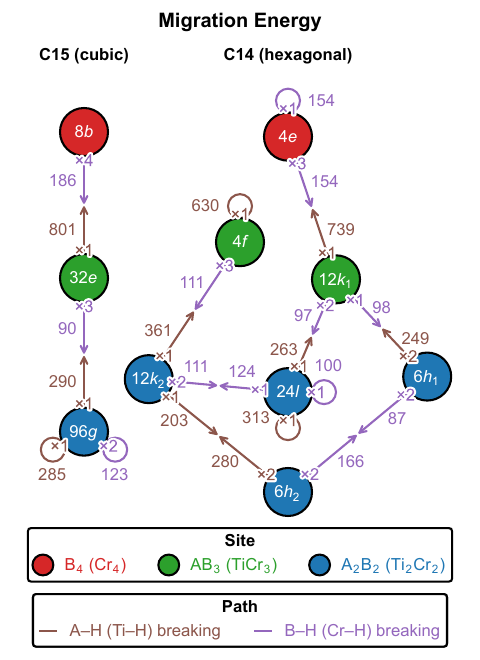}
\caption{All symmetrically distinct migration paths between face-sharing interstitial sites in the dilute limit (single hydrogen atom in the simulation cell). Arrows correspond to the available paths for hydrogen to migrate between two sites, along with their multiplicities and hydrogen migration barriers (meV) in Table~\ref{tab:migration}. Loops indicate migration between the sites of the same type.}
\label{fig:jumps}
\end{figure}

The C15 cubic and the C14 hexagonal Laves phases exhibit 4 and 11 symmetrically distinct first-nearest-neighbor (1NN) face-sharing interstitial pairs, respectively, between which hydrogen atoms may migrate.
Table~\ref{tab:migration} shows the migration energies of a single hydrogen atom for all the symmetrically distinct migration paths in the \ce{TiCr2} Laves phases obtained from DFT in conjunction with the G-SSNEB method.
Additionally, Fig.~\ref{fig:jumps} visually summarizes these migration paths, together with the migration energies.
\textcolor{responsered}{The minimum-energy paths for these migration paths are visualized in Sec.~\textcolor{violet}{S2} in the SM.
The migration energies obtained with the present MTPs are given in Sec.~\textcolor{violet}{S4} in the SM.}

The migration energy for each path is qualitatively characterized by the bond that needs to be broken during the migration; a Ti--H bond requires a higher energy to break than a Cr--H bond.
Specifically, the migration energies involving the breaking of a Ti--H bond span \qtyrange[color=responsered]{285}{801}{meV} and \qtyrange{203}{739}{meV} for the C15 and the C14 phases, respectively.
In contrast, the migration energies involving the breaking of a Cr--H bond span \qtyrange[color=responsered]{90}{186}{meV} and \qtyrange[color=responsered]{87}{154}{meV} for the C15 and the C14 phases, respectively.
This demonstrates higher chemical affinity of Ti for H than that of Cr for H.
Since A and B in other typical \ce{AB2} Laves-phase alloys also tend to show higher and lower chemical affinity with H, respectively, the trend derived above may also hold for these Laves-phase alloys, i.e., the migrations involving the breaking of an A--H bond also likely show higher migration barriers than the migrations involving the breaking of a B--H bond.

A distinction between the migrations involving the breaking of Ti--H and Cr--H bonds is particularly important for the C15 $96g$--$96g$ and the C14 $24l$--$24l$ paths.
For each of these pairs, there are two symmetrically distinct paths: one with the breaking of a Ti--H bond and the other with the breaking of a Cr--H bond.
The migration energies of the paths with the Ti--H breaking are \qtylist[color=responsered]{285;313}{meV} for the C15 $96g$--$96g$ and the C14 $24l$--$24l$ pairs, respectively, while the migration energies of the paths with the Cr--H breaking are \qtylist[color=responsered]{123;100}{meV} for the C15 $96g$--$96g$ and the C14 $24l$--$24l$ pairs, respectively.
This exemplifies the importance of considering not only the symmetry of each end point but also the symmetry of the path in between when evaluating the hydrogen migration between the interstitial sites.

The migrations from the \ce{AB3} to the \ce{B4} sites require much higher migration energies than the others;
the migration energy from the $32e$ to the $8b$ sites in the C15 phase is \qty[color=responsered]{801}{meV}, and the migration energy from the $4f$ to the $4e$ sites in the C14 phase is \qty{739}{meV}.
It is therefore unlikely that hydrogen atoms visit the \ce{B4} sites during diffusion.

\begin{figure}[!tbp]
\centering
\includegraphics{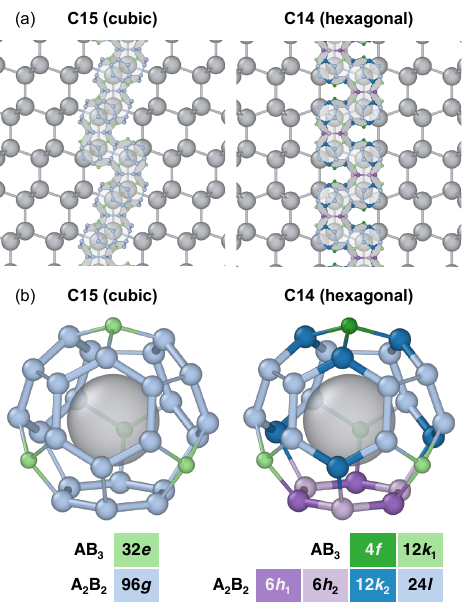}
\caption{Cages of \ce{A2B2} and \ce{AB3} interstitial sites surrounding A atoms in the \ce{AB2} Laves phases, which correspond to the diffusion pathways of hydrogen.
Colored spheres represent interstitial sites, where the \ce{A2B2} sites are shown as larger spheres compared to the \ce{AB3} sites.
Bonds between the interstitial-site spheres represent possible migration paths for hydrogen atoms, where thicker bonds show low-migration-energy paths within hexagonal rings associated with B–H bond breaking, and thinner bonds denote high-migration-energy paths requiring A–H bond breaking when jumping from the \ce{A2B2} sites.
In the C15 Laves phase, the 96\textit{g} sites form low-migration-energy hexagonal rings.
In the C14 Laves phase, the 6\textit{h}\textsubscript{1}–6\textit{h}\textsubscript{2} and the 12\textit{k}\textsubscript{2}-24\textit{l} sites form symmetrically distinct low-migration-energy hexagonal rings.
(a)~Part of the super-network of the cages in each Laves phase. (b)~Expansion of a single cage in each Laves phase.}
\label{fig:Cage_combined}
\end{figure}

The paths among the \ce{A2B2} and the \ce{AB3} interstitial sites in each Laves phase make a network, as visualized in Fig.~\ref{fig:Cage_combined}.
The network consists of cages, each of which surrounds an A atom and is topologically analogous to the \ce{C28} fullerene~\cite{kroto_stability_1987}.
The cages in each Laves phase further build a super-network in a similar manner as the A atoms, i.e., cubic and hexagonal diamond networks for the C15 and the C14 Laves phases, respectively.
Each cage contains four hexagonal rings consisting of \ce{A2B2} sites sharing \ce{A2B} faces.
A hydrogen jump among these sites involves the breaking of a B–H bond.
In contrast, a hydrogen jump escaping from a hexagonal ring involves the breaking of an A–H bond, which requires a higher energy than a B–H bond.
Therefore, the migration barriers of the inter-ring jumps are higher than those of the intra-ring jumps.

In the C14 hexagonal Laves phase, due to the symmetrical distinction among the \ce{A2B2} interstitial sites (cf.~Sec.~\ref{sec:methodology:models}), each cage has two distinct types of hexagonal rings.
Specifically, among the four hexagonal rings in each cage, one consists of 6\textit{h}\textsubscript{1} and 6\textit{h}\textsubscript{2} sites and the other three consist of 12\textit{k}\textsubscript{2} and 24\textit{l} sites.
In the case of C14 \ce{TiCr2} investigated in the present study, among the inter-ring jumps, the 6\textit{h}\textsubscript{2}-to-6\textit{h}\textsubscript{1} migration shows the highest migration barrier (\qty[color=responsered]{166}{meV}), followed by 24\textit{l}-to-12\textit{k}\textsubscript{2} (\qty[color=responsered]{124}{meV}), 12\textit{k}\textsubscript{2}-to-24\textit{l} (\qty{111}{meV}), 24\textit{l}-to-24\textit{l} (\qty[color=responsered]{100}{meV}), and 6\textit{h}\textsubscript{1}-to-6\textit{h}\textsubscript{2} (\qty[color=responsered]{87}{meV}) jumps in descending order (cf.~Sec.~\ref{sec:results:migration_energies}).
This indicates that, within the hexagonal rings, hydrogen atoms at the 6\textit{h}\textsubscript{2} sites may take the longest time to jump, as also expected from the most negative binding energies at the 6\textit{h}\textsubscript{2} site (cf.~Sec.~\ref{sec:results:solution_energies}).
In consequence, the characteristics of hydrogen diffusion are more sophisticated in the C14 hexagonal \ce{TiCr2} than in the C15 cubic counterpart (cf.~\ref{sec:anisotropy}).
The detailed impact of the symmetrically distinct sites in the C14 Laves phase can vary with its chemical composition, and hence we may find different trends for other Laves-phase alloys.

\subsection{Mean squared displacements of hydrogen}
\label{sec:results:MSD}

To investigate hydrogen diffusion at finite hydrogen concentrations, we performed MD simulations using the C15-MTP and the fine-tuned C14-MTP (cf.~Sec.~\ref{sec:methodology:MLIPs}).

\begin{figure}[tbp]
\centering
\includegraphics{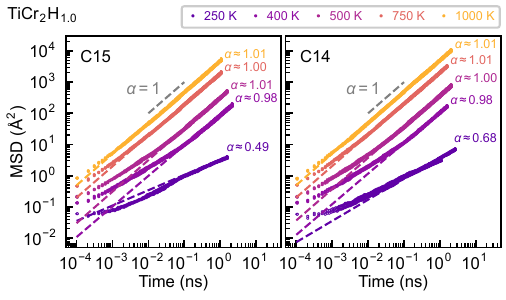}
\caption{MSDs of hydrogen atoms in \ce{TiCr2H_{1.0}} as a function of MD simulation time.}
\label{fig:MSD_16H_all}
\end{figure}

We first examined if the hydrogen diffusion reaches the normal-diffusion regime via Eq.~\eqref{eq:msd_alpha}.
Figure~\ref{fig:MSD_16H_all} illustrates the evolution of the MSD in \ce{TiCr2H_{1.0}} over MD time at five different temperatures: \qtylist{250;400;500;750;1000}{K}.
At and above \qty{400}{K}, as time progresses, the slope of the MSD with respect to MD time in the log–log scale, i.e., $\alpha$ in Eq.~\eqref{eq:msd_alpha}, converges to 1, indicating that the MD simulations reach the normal-diffusion regime where the diffusion coefficients can be evaluated via Eq.~\eqref{eq:D_alpha}. In contrast, the MSDs at \qty{250}{K} exhibit reduced slopes and fail to reach the normal-diffusion regime within the nanosecond time scale accessible in MD simulations.
The same trend is found for the other hydrogen concentrations (cf. Sec. \textcolor{responsered}{S4} in the \textcolor{violet}{SM}).

\begin{figure}[!tbp]
\centering
\includegraphics{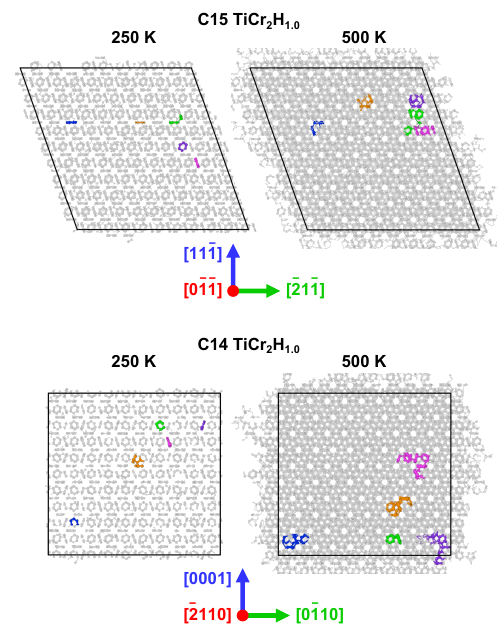}
\caption{Trajectories of hydrogen atoms in C15 and C14 \ce{TiCr2H_{1.0}} during MD simulations. Gray lines represent the trajectories of all hydrogen atoms over \qty{0.1}{ns} of simulation, while thick colored lines highlight the individual trajectories of five randomly selected hydrogen atoms.}
\label{fig:trajectories}
\end{figure}

Figure~\ref{fig:trajectories} shows the trajectories of hydrogen atoms in \ce{TiCr2H_{1.0}} at both \qty{250}{K} and \qty{500}{K} after an MD simulation time of \qty{0.1}{ns}.
At \qty{250}{K}, almost all hydrogen atoms remain confined within the hexagonal rings made of \ce{A2B2} sites due to the respective lower migration barriers, rarely migrating to neighboring hexagonal rings.
Inter-ring migrations are necessary for hydrogen to reach the normal-diffusion regime, but their occurrence is statistically insufficient at this low temperature even after \qty{1}{ns} of MD simulation time (Fig.~\ref{fig:MSD_16H_all}).
In contrast, at \qty{500}{K}, most hydrogen atoms undergo multiple inter-ring migrations, thereby clearly exhibiting long-range diffusion. This confirms that the system is in the normal-diffusion regime for hydrogen.

\subsection{Dependence on hydrogen concentration}
\label{sec:results:D_vs_x}

\begin{figure}[!tb]
\centering
\includegraphics{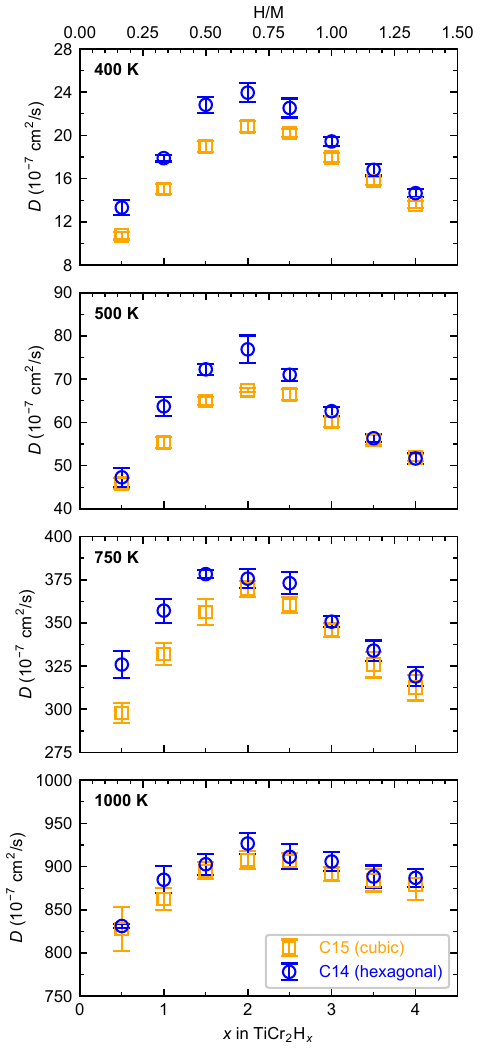}
\caption{Diffusion coefficients of hydrogen atoms as a function of hydrogen concentration in TiCr\textsubscript{2} at four temperatures. Error bars represent the standard deviations obtained from five independent MD simulations.}
\label{fig:HM_Dif}
\end{figure}

\begin{figure}[tb]
\centering
\includegraphics{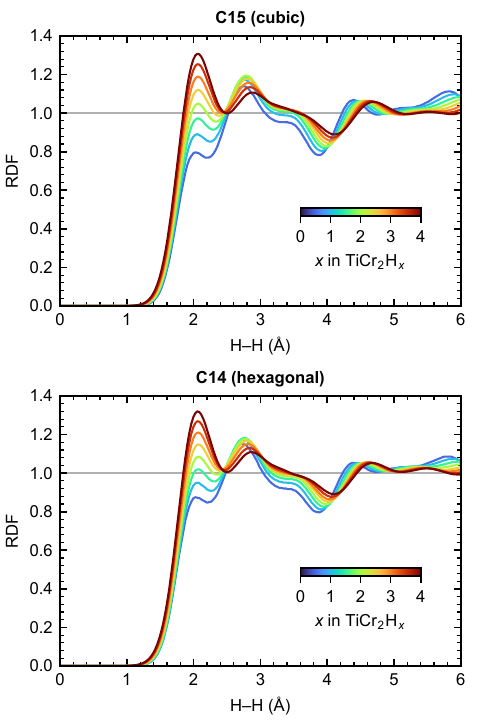}
\caption{Time averaged RDFs for the H--H pairs in \ce{TiCr2H_x} obtained with the MD simulations at \qty{1000}{K}.}
\label{fig:RDFs}
\end{figure}

We next discuss the dependence of hydrogen diffusion coefficients on hydrogen concentration.
As detailed in Sec.~\ref{sec:results:MSD}, at \qty{250}{K}, the systems did not reach the normal-diffusion regime within the MD simulation time, and therefore we could not obtain the diffusion coefficients via Eq.~\eqref{eq:D_alpha}.
Note also that quantum effects could be substantial at such low temperatures (cf.~Sec.~\ref{sec:methodology:diffusion}).
We have therefore computed diffusion coefficients only at higher temperatures.
Figure~\ref{fig:HM_Dif} presents the diffusion coefficients \textit{D} of hydrogen at four temperatures as a function of $x$ in \ce{TiCr2H_x}.
(The raw values of $D$ are provided in the \textcolor{violet}{SM}.)

For each temperature and each hydrogen concentration, the diffusion coefficients tend to be lower in the C15 cubic phase than the C14 hexagonal phase, but only slightly.
This reflects the geometrical similarity of the two Laves phases, where topologically only the stacking sequences are different \cite{gieselmann_laves_2023}.

For all temperatures, the diffusion coefficient $D$ substantially depends on the hydrogen concentration \textit{x}. This dependence is non-monotonic; $D$ increases for $0 < x \le 2$ and decreases for $2 \le x \le 4$ with increasing \textit{x}.
The dependence on hydrogen concentration is more evident at lower temperatures.
At the highest considered temperature of \qty{1000}{K}, the diffusion coefficient increases modestly by 13\% in the C15 phase and 10\% in the C14 phase when $x$ increases from 0.5 to 2.0.
At \qty{400}{K}, the increases are much larger, accounting for 94\% and 80\% in the C15 and the C14 phases, respectively, within the same $x$ range.

To investigate the trend of hydrogen occupation at the interstitial sites and its impact on diffusion coefficients, we also calculated the time-averaged radial distribution functions (RDFs) of hydrogen atoms to analyze their occupancy trends at \qty{1000}{K}, as visualized in Figure~\ref{fig:RDFs}.
The time-averaging was conducted over the last \qty{50}{ps} of one of the five MD trajectories used for the diffusion-coefficient calculations.

For each hydrogen concentration, the time-averaged RDFs reveal two prominent peaks: one near \qty{2.1}{\AA}, corresponding to the Switendick limit for H--H separation, and another around \qtyrange{2.7}{2.8}{\AA}.
There are no peaks around \qty{1.6}{\AA}, i.e., the distances between the 1NN face-sharing interstitial sites~\cite{shoemaker_concerning_1979}.
This indicates that the face-sharing positions are rarely occupied during hydrogen diffusion, likely due to strong repulsion between hydrogen atoms at these positions~\cite{KUMAR2025121319}.
Therefore, the first peak near \qty{2.1}{\AA} primarily arises from hydrogen atoms in the second-nearest-neighbor (2NN) edge-sharing interstitial sites, while the second peak reflects occupancy at more distant sites.

At a low hydrogen concentration of $x = 0.5$, hydrogen atoms tend to occupy isolated positions, as evidenced by the lower height of the first peak in the RDFs compared to the second peak.
As the hydrogen concentration increases, the height of the first peak grows progressively, while the second peak remains nearly saturated.
This enhancement of the first peak indicates that hydrogen atoms increasingly occupy the 2NN edge-sharing interstitial sites, resulting in stronger repulsion between hydrogen atoms.
This repulsive interaction influences hydrogen diffusion in two regimes.
For $x < 2$, the growing collective repulsion acts as an internal driving force that lowers the effective activation barrier for hydrogen migration between interstitial sites, thereby enhancing diffusion.
Around $x \approx 2$, the first RDF peak reaches a height comparable to the height of the second peak,
indicating that the local hydrogen concentrations at the edge-sharing positions become nearly equal to the average hydrogen concentrations in the systems.
The diffusion coefficients become highest around this concentration, as found in Fig.~\ref{fig:HM_Dif}.
For $x > 2$, the continued increase of the first peak reflects excessive crowding, where intense repulsion between hydrogen atoms restricts hydrogen migration.
That is, an unoccupied interstitial site with more than one adjacent site occupied by hydrogen atoms is blocked so that one of these hydrogen atoms cannot migrate into the unoccupied site due to the strong repulsion with the other hydrogen atoms at the face-sharing positions.
This eventually leads to the declines in diffusion coefficients.

\subsection{Comparison with experiments}

\begin{figure}[tb]
\centering
\includegraphics{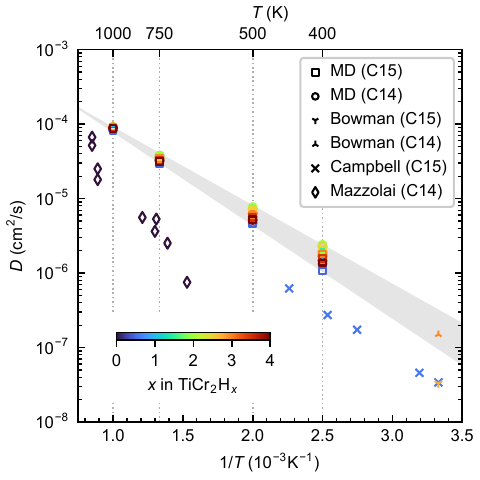}
\caption{Diffusion coefficients of hydrogen atoms in \ce{TiCr2H_x} in the C15 and the C14 Laves phases as a function of inverse temperature, calculated through MD simulations.
The colors of the points represent the hydrogen concentration \textit{x} as specified by the color scale.
The shaded region indicates the range of diffusion coefficients for varying hydrogen concentrations.
Experimental values for C15 \ce{TiCr_{1.8}H_{2.58}} and C14 \ce{TiCr_{1.9}H_{2.85}} reported by Bowman \textit{et al.}~\cite{bowman_diffusion_1983}, C15 \ce{TiCr_{1.85}H_{0.43}} reported by Campbell \textit{et al.}~\cite{campbell_quasi-elastic_1999}, and C14 \ce{TiCr_{1.85}H_{0.017}} reported by Mazzolai \textit{et al.}~\cite{mazzolai_hydrogen_2009} are also included for comparison.}
\label{fig:Dif}
\end{figure}

Figure \ref{fig:Dif} shows the computed hydrogen diffusion coefficients of \ce{TiCr2H_x} as a function of inverse temperature across different hydrogen concentrations.
The MD simulations reveal that, across the entire investigated temperature range above \qty{400}{K}, the diffusion coefficients at each hydrogen concentration show an Arrhenius-type dependence.
As discussed in Sec.~\ref{sec:results:D_vs_x}, at a fixed temperature, the diffusion coefficients show a non-monotonic dependence on hydrogen concentration.
Thus, the diffusion coefficients in Fig.~\ref{fig:Dif} are distributed in the shaded region, and the corresponding activation barriers lie in the range \qtyrange{0.21}{0.25}{eV}.
These activation barriers are in good agreement with many experimental values.
Specifically, Hiebl~\cite{hiebl_proton_1982} obtained \qtyrange{0.2}{0.3}{eV} with proton and deuteron NMR measurements for C15 \ce{TiCr_{1.85}H_{0.8}} above \qty{250}{K}.
Campbell \textit{et al.}~\cite{campbell_quasi-elastic_1999} obtained \qty{0.24}{eV} with quasi-elastic neutron scattering for C15 \ce{TiCr_{1.85}H_{0.43}} at \qtyrange{313}{442}{K}.
Bowman~\textit{et al.}~\cite{bowman_diffusion_1983} obtained \qtyrange{0.13}{0.26}{eV} and \qtyrange{0.205}{0.40}{eV} for C15 \ce{TiCr_{1.8}H_{x}} (\textit{x}~= 0.55, 2.58) and C14 \ce{TiCr_{1.9}H_{x}} (\textit{x}~= 0.63, 2.85), respectively, above \qty{180}{K}.
In contrast, Mazzolai \textit{et al.}~\cite{mazzolai_hydrogen_2009} reported an activation energy of approximately \qty{0.6}{eV} for C14 \ce{TiCr_{1.78}H_{0.017}} at temperatures at \qtyrange{660}{1200}{K}, determined from pressure–time measurements for hydrogen absorption.
This high value was interpreted as evidence of classical over-barrier hopping, in contrast to the low- and intermediate-temperature hydrogen dynamics probed by NMR, QENS, or internal friction.
However, our present MD results actually show an Arrhenius-type trend over a wide temperature range of \qtyrange{400}{1000}{K}.
We would therefore ascribe the exceptionally high activation barrier in Ref.~\cite{mazzolai_hydrogen_2009} rather to difficulties of the respective experimental approach in determining the intrinsic hydrogen diffusion coefficients.
In further support, we note that other binary Laves phases also show activation barriers in the range \qtyrange{0.13}{0.26}{eV} in experiments~\cite{skripov_hydrogen_2003}, consistently with the present MD simulations.

Despite the agreement of the computed activation barrier with the experimental values, the MD simulations show diffusion coefficients one order of magnitude higher than the experimental values, particularly those of Campbell~\textit{et al.}~\cite{campbell_quasi-elastic_1999}.
The most likely reason for the discrepancy is the non-stoichiometric compositions of metal atoms in experiments, i.e., \ce{TiCr_{1.8}} for the C15 phase~\cite{johnson_reaction_1978} and \ce{TiCr_{1.9}} for the C14 phase~\cite{johnson_reaction_1980}, which are not considered in the present MD simulations.
The non-stoichiometric compositions require either Cr vacancies or Ti anti-sites at the Cr sites.
Indeed, vacancies in metallic systems can act as hydrogen traps~\cite{zhong_hydrogen_2025}.
Also, anti-site Ti indicates a higher ratio of the \ce{A2B2} (\ce{Ti2Cr2}) interstitial sites than those in the stoichiometric composition, which show larger binding energies than the other types of interstitial sites (cf.~Sec.~\ref{sec:results:solution_energies}) and thus possibly slow down hydrogen kinetics.
Intrinsic errors in the exchange--correlation functional in DFT can be another reason~\cite{kimizuka_mechanism_2018}.
Specifically, the PBE functional tends to underestimate the binding energies among atoms \cite{grabowski_ab_2007}, which may result in overestimation of diffusion coefficients.
Note also that the experimental values of Bowman~\textit{et al.}~\cite{bowman_diffusion_1983} show diffusion coefficients that differ by one order of magnitude for C15 \ce{TiCr_{1.8}H_{2.58}} (\qty{3.4e-8}{cm^2/s}) and C14 \ce{TiCr_{1.9}H_{2.85}} (\qty{1.5e-7}{cm^2/s}) at \qty{300}{K}. While the former agrees well with the values of Campbell \textit{et al.}, the latter is within the range expected from the present MD simulations based on the Arrhenius relationship extrapolated to \qty{300}{K}.

\section{Conclusions} 
\label{sec:Conclusions}

In this work, we have elucidated the mechanisms of hydrogen diffusion in \ce{TiCr2H_x} Laves phases ($0 < x \le 4$) and revealed that complex H–H interactions lead to changes in diffusion coefficients with hydrogen concentration.

The DFT-calculated migration barriers in the dilute hydrogen limit exhibit substantially different hydrogen migration energies even among the \ce{A2B2} sites. Specifically, the activation barriers for the paths involving the breaking of a \ce{Ti-H} bond are substantially higher than those involving the breaking of a \ce{Cr-H} bond, demonstrating higher chemical affinity of Ti for H than that of Cr.

The MD simulations across varying hydrogen concentrations reveal a non-monotonic dependence of diffusion coefficients on hydrogen content, which is more pronounced at lower temperatures.
Time-averaged RDFs of hydrogen atoms indicate that hydrogen atoms avoid occupying 1NN face-sharing positions even during diffusion.
As the hydrogen concentration increases, more 2NN edge-sharing interstitial sites get occupied, leading to an increase in H--H repulsion.
This repulsive interaction influences hydrogen diffusion in two regimes.
At lower hydrogen concentrations (\textit{x} $<$ 2), the repulsion interaction acts as a driving force to reduce the activation barrier.
At higher hydrogen concentrations (\textit{x} $>$ 2), the repulsion blocks the remaining interstitial sites from being occupied, hindering hydrogen migration.

The MD-derived diffusion coefficients exhibit an Arrhenius-type behavior in the investigated temperature range (\qtyrange{400}{1000}{K}), with activation barriers in good agreement with most experimental values.
In contrast, the diffusion coefficients are approximately one order of magnitude higher than most experimental values.
The discrepancy is likely due to hydrogen trapping by defects such as Cr vacancies and Ti anti-sites in the non-stoichiometric \ce{TiCr2} compositions in experiments, not captured in the present MD simulations.

The above findings provide various atomistic insights into hydrogen diffusion in \ce{TiCr2}.
The gained knowledge may also be transferable to multi-component Laves phases for more practical hydrogen storage applications, providing guidance to tailor hydrogen kinetics for fast hydrogen absorption and desorption.

\section*{CRediT authorship contributions}

\textbf{Pranav Kumar}: Conceptualization, Methodology, Software, Validation, Formal Analysis, Investigation, Data Curation, Writing – Original Draft, Visualization.
\textbf{Fritz Körmann}: Writing – Review \& Editing.
\textbf{Kaveh Edalati}: Writing – Review \& Editing.
\textbf{Blazej Grabowski}: Resources, Funding Acquisition, Writing – Review \& Editing.
\textbf{Yuji Ikeda}: Conceptualization, Software, Visualization, Writing – Review \& Editing, Supervision, Project Administration, Funding Acquisition.

\section*{Declaration of Generative AI and AI-assisted technologies in the writing process}

ChatGPT and Grammarly were used to assist with initial drafting and sentence refinement for certain sections of this manuscript. The authors subsequently reviewed and edited all content as needed and take full responsibility for the final version of the publication.

\section*{Declaration of competing interest}

The authors declare that they have no known competing financial interests or personal relationships that could have appeared to influence the work reported in this paper.

\section*{Acknowledgments}

Pranav Kumar and Yuji Ikeda are funded by the Deutsche Forschungsgemeinschaft (DFG, German Research Foundation), Project No.~519607530.
Fritz Körmann acknowledges support by the Heisenberg Programme of the Deutsche Forschungsgemeinschaft (DFG, German Research Foundation), Project No.~541649719.
Blazej Grabowski acknowledges funding from the European Research Council (ERC) under the European Union’s Horizon 2020 research and innovation programme (grant agreement No.~865855).
The authors also acknowledge support by the state of Baden--Württemberg through bwHPC
and the DFG through grant no INST 40/575-1 FUGG (JUSTUS 2 cluster) and the SFB1333 (project ID 358283783-CRC 1333/2 2022).

\appendix

\setcounter{table}{0}
\setcounter{figure}{0}

\makeatletter
\gdef\thesection{\appendixname\@Alph\c@section}
\makeatother

\section{Impact of supercell size on diffusion coefficients}
\label{sec:D_vs_supercell}

\begin{figure}[tb]
\centering
\includegraphics[scale=1]{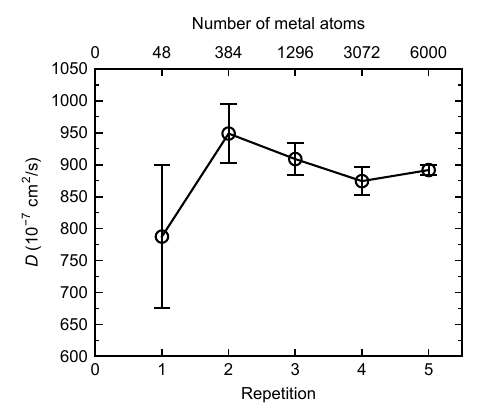}
\caption{Diffusion coefficient of C15 TiCr$_2$H$_{3.0}$ as a function of supercell size. Error bars represent the standard deviations obtained from five independent MD simulations.}
\label{fig:periodicity}
\end{figure}

In this appendix, we investigate the effect of supercell size on the diffusion coefficients of hydrogen, taking C15 \ce{TiCr2H_{3.0}} as an example.
Specifically, we examined further $n \times n \times n$ (\textit{n} = 1, 2, 3, 4, 5) expansions of the original 48-metal-atom cells (cf.~Sec.~\ref{sec:methodology:models}). MD simulations were performed at \qty{1000}{K}. The system was initially equilibrated for \qty{50}{ps} in the \textit{NPT} ensemble, followed by the calculation of the MSD for \qty{1}{ns} in the \textit{NVE} ensemble.
Figure~\ref{fig:periodicity} shows the thus obtained diffusion coefficients of hydrogen in C15 \ce{TiCr2H_{3.0}} as a function of supercell size.
As the supercell size increases, the number of hydrogen atoms also increases, providing better statistical sampling and leading to diffusion coefficients with reduced uncertainty.
With 6000 metal atoms, i.e., for the \numproduct{5x5x5} repetition of the original 48-metal-atom cells, the uncertainty of the diffusion coefficient is less than \qty{1}{\percent}.
We therefore use the 6000-metal-atom cells for the investigation of diffusion coefficients in the present study.

\section{Anisotropy of hydrogen diffusion 
in C14 TiCr\textsubscript{2}}
\label{sec:anisotropy}

\begin{figure*}[!bt]
\centering
\includegraphics{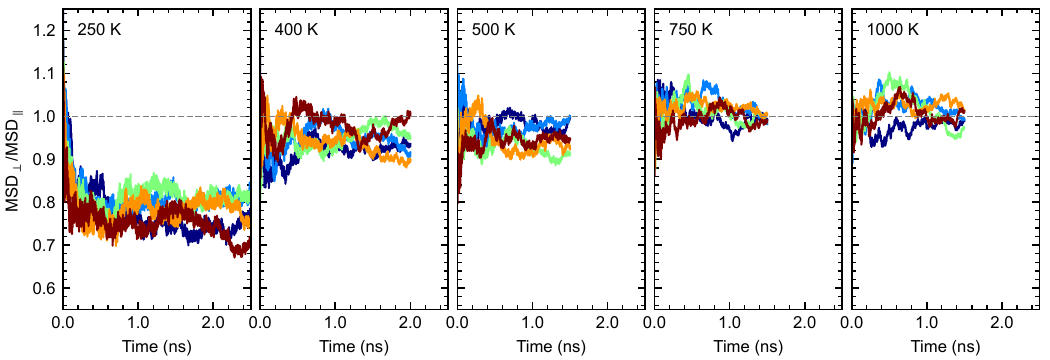}
\caption{Ratio of MSDs of hydrogen atoms parallel and perpendicular to the [0001] direction in C14 \ce{TiCr2H_{1.0}} from five independent MD simulations at different temperatures.}
\label{fig:anisotropy}
\end{figure*}

Since the C14 Laves phase has hexagonal symmetry, in principle the hydrogen diffusion coefficients are isotropic and should show differences between the directions perpendicular and parallel to the [0001] direction.
In this appendix, we investigate the degree of anisotropy of hydrogen diffusion via the hydrogen MSDs resolved into the contribution of a direction perpendicular to [0001]
\begin{equation}
\text{MSD}_\perp(t) = \frac{1}{N_\mathrm{H}} \sum_{i=1}^{N_\mathrm{H}} \frac{1}{2}
\Big(
  \left| x_i(t) - x_i(0) \right|^2
+ \left| y_i(t) - y_i(0) \right|^2
\Big)
\end{equation}
and that along [0001]
\begin{equation}
\text{MSD}_\parallel(t) = \frac{1}{N_\mathrm{H}} \sum_{i=1}^{N_\mathrm{H}} \left| z_i(t) - z_i(0) \right|^2,
\end{equation}
where the \textit{z} axis is set along the [0001] direction.
Five MD simulations, independent of those for the main-text results, were run for \ce{TiCr2H_{1.0}} at temperatures of \qtylist{250;400;500;750;1000}{K} under the same MD conditions as those in the main text.
Figure~\ref{fig:anisotropy} presents the MSD$_\parallel$/MSD$_\perp$ ratios obtained from the MD simulations.

At \qty{250}{K}, where hydrogen does not reach the normal-diffusion regime within an MD simulation time of \qty{2.5}{ns} (cf.~Sec.~\ref{sec:results:MSD}),
MSD$_\parallel$ is substantially smaller than MSD$_\perp$, with the MSD$_\parallel$/MSD$_\perp$ ratios in a range of 0.7–0.8.
This indicates that, in the given short time scale, hydrogen tends to be more trapped along the [0001] direction compared with the directions perpendicular to [0001] (cf.~Sec.~\ref{sec:results:MSD}).
This happens maybe because more hydrogen are trapped in the 6\textit{h}\textsubscript{1}–6\textit{h}\textsubscript{2} rings than the 12\textit{k}\textsubscript{2}-24\textit{l} rings (cf.~Sec.~\ref{sec:results:migration_energies}).
That is, the 6\textit{h}\textsubscript{2} sites show substantially more negative hydrogen binding energies than the other \ce{A2B2} sites (cf.~Sec.~\ref{sec:results:solution_energies}) and show the highest migration barriers than the other paths within the hexagonal rings (cf.~Sec.~\ref{sec:results:migration_energies}).
Since the 6\textit{h}\textsubscript{1}–6\textit{h}\textsubscript{2} rings lie perpendicular to the [0001] direction, they does not contribute to MSD$_\parallel$.

At temperatures of \qty{400}{K} and above—where the MD simulations reach the normal-diffusion regime—the ratio of MSD$_\parallel$ to MSD$_\perp$ converges to values between 0.9 and 1.1.
This suggests that, in contrast to the short length scale, long-range hydrogen diffusion in C14 \ce{TiCr2} is essentially isotropic at the higher temperatures.

\section*{Data availability}

The developed MTPs along with the corresponding DFT training datasets
are freely accessible on \href{https://darus.uni-stuttgart.de/previewurl.xhtml?token=50ff2068-cbb1-4df8-b785-60cf623115bc}{DaRUS}.

\bibliographystyle{elsarticle-num} 

\end{document}


\begin{frontmatter}
\title{Hydrogen diffusion in \texorpdfstring{TiCr\textsubscript{2}H\textsubscript{\textit{x}}}{TiCr2Hx} Laves phases: A combined \textit{ab initio} and machine-learning-potential study\\\texorpdfstring{\vspace{\baselineskip}}{}Supplementary Materials}
\end{frontmatter}

\section{Impact of supercell size on the binding and the migration energies of a single hydrogen atom}

To evaluate the impact of supercell size in our DFT calculations, we examined the convergence of the binding and the migration energies of a single hydrogen atom in \ce{TiCr2} with respect to supercell size. We considered a single hydrogen atom in the C15 Laves phase. The primitive cell of the C15 phase contains six metal atoms, so that $1\times1\times1$, $2\times2\times2$, $3\times3\times3$, and $4\times4\times4$ expansions of the primitive cell correspond to 6, 48, 162, and 384 metal atoms, respectively.
The hydrogen binding energy at the $96g$ (A$_2$B$_2$) site was computed for all these supercell sizes.
The migration energy of the $96g$--$96g$ path involving B--H bond breaking, which shows a lower migration barrier than that involving A--H bond breaking, was evaluated up to the $3\times3\times3$ supercell.
We employed 11 intermediate images for the $1\times1\times1$ and $2\times2\times2$ supercells to ensure sufficient resolution of the migration path, while for the $3\times3\times3$ supercell, three intermediate images were used due to the substantially higher computational cost.

\begin{figure*}[htbp]
\centering
\includegraphics{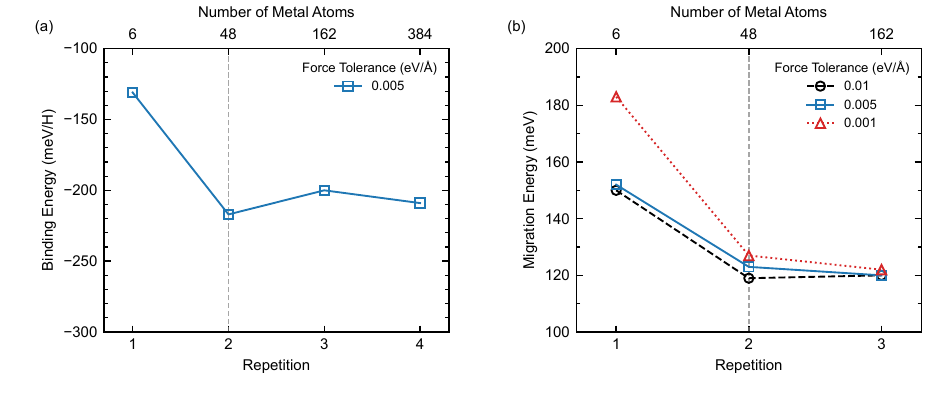}
\caption{Convergence of hydrogen binding and migration energies in the C15 Laves phase of \ce{TiCr2} with respect to supercell size.
Vertical dashed lines indicate the \numproduct{2x2x2} supercell used in the present manuscript.
(a) Hydrogen binding energy at the $96g$ site.
(b) Hydrogen migration energy for the $96g$–$96g$ path involving B--H bond breaking calculated using the G-SSNEB method with force convergence tolerances of 0.01, 0.005, and 0.001~eV/\AA.}
\label{fig:cell_conv_DFT}
\end{figure*}

Figure~\ref{fig:cell_conv_DFT} presents the thus obtained binding and migration energies.
Both the binding and the migration energies of hydrogen are well converged for the \numproduct{2x2x2} supercell and above.
The difference in binding energy between the $2\times2\times2$ and the $3\times3\times3$ supercells is less than \qty{20}{meV}, while the difference in migration energy is below \qty{5}{meV} for these supercells. These results suggest that the $2\times2\times2$ supercell offers a good balance between computational efficiency and accuracy.

In Fig.~\ref{fig:cell_conv_DFT}, we also show the impact of the force convergence tolerance on the migration energy obtained from the G-SSNEB method.
For the \numproduct{2x2x2} supercell, force tolerances of 0.01, 0.005, and 0.001~eV/\AA\ result in migration energies of \qtylist{119;123;127}{meV}, respectively. The difference between 0.01 and 0.005~eV/\AA\ is less than \qty{5}{meV}, and increasing the supercell to $3\times3\times3$ changes the barrier by only \qty{3}{meV} for 0.005~eV/\AA.
Therefore, all G-SSNEB calculations in the main text were performed using a force tolerance of \qty{0.005}{eV/\AA}.

\section{Minimum-energy paths}
\label{sec:NEB}

\begin{figure*}[tbp]
\centering
\includegraphics{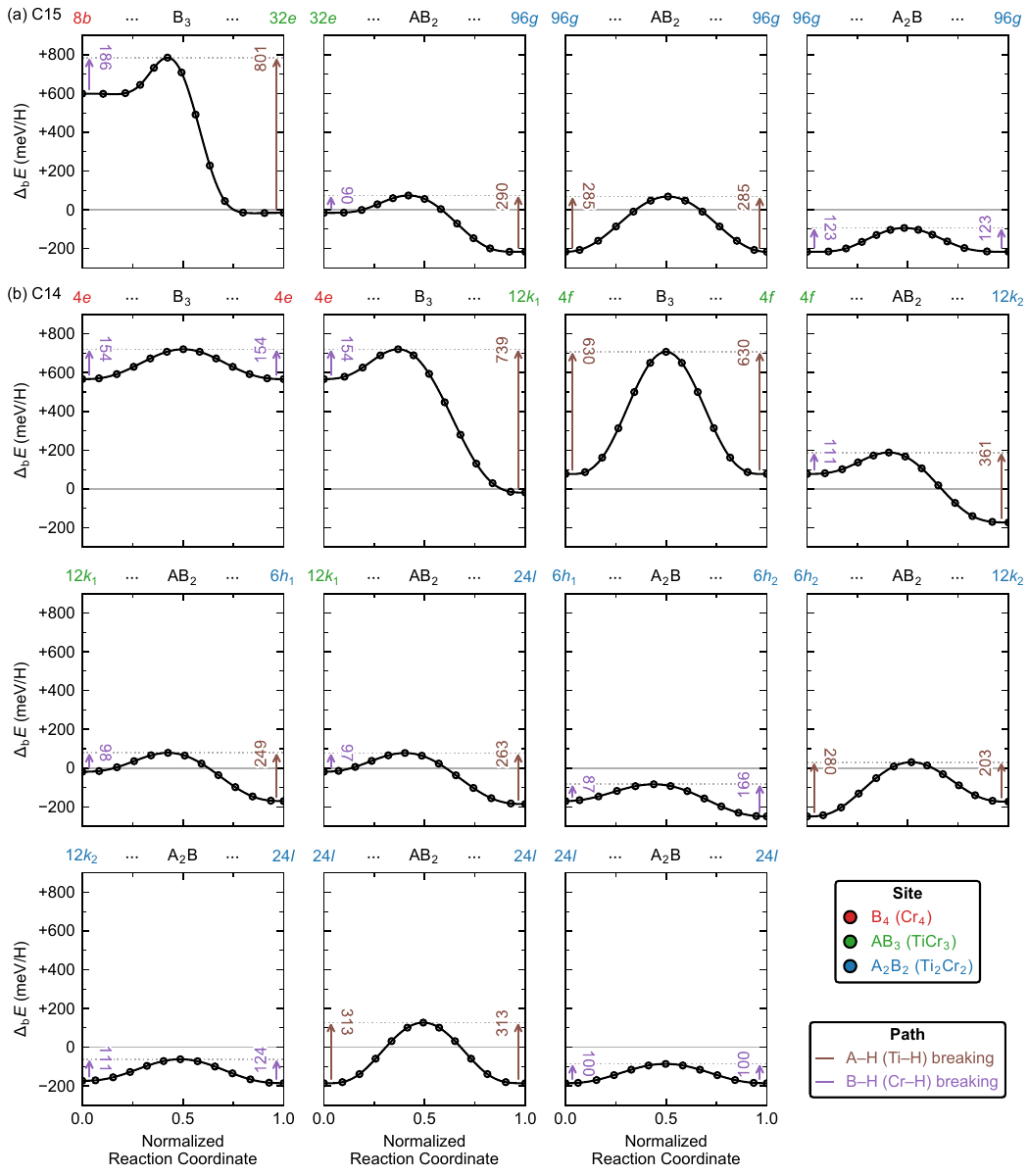}
\caption{Minimum-energy paths for all symmetrically inequivalent paths of \ce{TiCr2} in the (a) C15 cubic and the (b) C14 hexagonal Laves phases obtained with DFT in conjunction with the G-SSNEB method.}
\label{fig:NEB}
\end{figure*}

Figure~\ref{fig:NEB} presents the minimum-energy paths for all symmetrically inequivalent paths of \ce{TiCr2} in the C15 cubic and the C14 hexagonal Laves phases obtained with DFT in conjunction with the G-SSNEB method.

\section{Comparison of the binding energies of a single hydrogen atom between DFT and MTP}

\begin{figure*}[!htbp]
\centering
\includegraphics{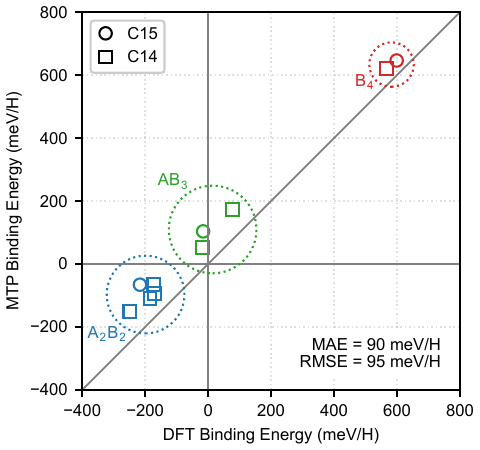}
\caption{
Comparison of DFT and MTP binding energies for different site types in the C14 and C15 phases. The mean absolute error (MAE) and root mean square error (RMSE) of the MTP values are displayed in the bottom-right corner of the plot.}
\label{fig:DFT_MTP_parity}
\end{figure*}

Table~\ref{tab:DFT_vs_MTP} and Fig.~\ref{fig:DFT_MTP_parity} show the binding energies of a single hydrogen atom in the \ce{TiCr2} Laves phases predicted by the MTPs refined in the present study and compares them with the corresponding DFT results (Table~\textcolor{violet}{2} in the main text). In both cases, the calculations were performed using supercells comprising 48 metal atoms, with simultaneous relaxation of atomic positions and lattice parameters.

The updated MTPs faithfully reproduce the energetic ordering obtained from DFT calculations. For both the C15 cubic and C14 hexagonal Laves phases, the \ce{A2B2} interstitial sites are predicted to be the most favorable for hydrogen incorporation, followed by the \ce{AB3} and \ce{B4} sites in descending order. In the C14 hexagonal phase, the $6h_2$ \ce{A2B2} interstices exhibit the lowest binding energies, while the $24l$ sites represent the next most stable hydrogen locations.

Note that the errors in the hydrogen binding energies are scaled by the total number of atoms in the supercells;
the RMSE of the MTP prediction for the binding energies of a single hydrogen atom is \qty{95}{meV/H}, while the RMSEs of the MTPs with respect to the training datasets are as small as \qty{3}{meV/atom} (Sec.~\textcolor{violet}{3.2} in the main text).
This highlights a general challenge for MLIPs in reproducing supercell-requiring defect properties with high accuracy.

\begin{table}[htb]
\centering
\small
\caption{Binding energies $E_\mathrm{b}$ of a single hydrogen atom at the interstitial sites of the C15 cubic and C14 hexagonal \ce{TiCr2} Laves phases, as obtained from DFT and MTP calculations.}
\label{tab:DFT_vs_MTP}
\begin{tabular}{cccS[table-format=+1.0]S[table-format=+1.0]}
\toprule
Phase
& Type
& Site
& {DFT (\unit{meV/H})}
& {MTP (\unit{meV/H})} \\
\midrule
C15
& \ce{B4}   & $8b$   &  599 &  647 \\
\cmidrule{2-5}
& \ce{AB3}  & $32e$  &  -16 &  104 \\
\cmidrule{2-5}
& \ce{A2B2} & $96g$  & -217 &  -66 \\
\midrule
C14
& \ce{B4}   & $4e$   &  566 &  621 \\
\cmidrule{2-5}
& \ce{AB3}  & $4f$   &   77 &  173 \\
&           & $12k_1$&  -19 &   53 \\
\cmidrule{2-5}
& \ce{A2B2} & $6h_1$ & -170 &  -94 \\
&           & $6h_2$ & -249 & -151 \\
&           & $12k_2$& -173 &  -65 \\
&           & $24l$  & -185 & -109 \\
\bottomrule
\end{tabular}
\end{table}

\section{Comparison of G-SSNEB migration energies from DFT and MTP}

Table \ref{tab_migration_MTP} presents a comparison of the G-SSNEB migration energies obtained from the trained MTPs alongside the corresponding DFT values (Table~\textcolor{violet}{3} in the main text).
The G-SSNEB calculations were performed within ASE~\cite{larsen_atomic_2017} using the TSASE library (\url{https://theory.cm.utexas.edu/tsase/}), employing the same G-SSNEB settings as for DFT calculations described in Sec.~\textcolor{violet}{3.2} in the main text.

The MTPs well reproduce the trend obtained from DFT.
Particularly, a Ti–H bond requires a higher energy to break than a Cr–H bond.
Specifically, the migration energies involving the breaking of a Ti--H bond span \qtyrange{293}{771}{meV} and \qtyrange{219}{711}{meV} for the C15 and the C14 phases, respectively.
In contrast, the migration energies involving the breaking of a Cr--H bond span \qtyrange{123}{206}{meV} and \qtyrange{100}{189}{meV} for the C15 and the C14 phases, respectively.
The RMSEs of the MTP prediction are \qty{26}{meV} and \qty{29}{meV} for the C15 and the C14 phases, respectively.

\color{responsered}

\begin{table*}[!htbp]
\centering
\small
\caption{Hydrogen migration energies for all symmetry-distinct diffusion paths in the \ce{TiCr2} Laves phases obtained with the MTPs.
The DFT values are also shown in parentheses for comparison.
The “Face” column indicates the common face between two interstitial sites through which hydrogen migrates.
The “Bond breaking” columns specify the types of metal–hydrogen bond broken during migration.}
\label{tab_migration_MTP}
\begin{tabular}{c c@{ }c c@{ }c c c c r@{ }r r@{ }r}
\toprule
&
\multicolumn{2}{c}{\multirow{2}{*}{\textit{S}\textsubscript{1}}} &
\multicolumn{2}{c}{\multirow{2}{*}{\textit{S}\textsubscript{2}}} &
\multirow{2}{*}{Face} &
\multicolumn{2}{c}{Bond breaking} &
\multicolumn{4}{c}{$E_\mathrm{mig}$ (meV)} \\
\cmidrule(lr){7-8}
\cmidrule(lr){9-12}
&&&&&&
\multicolumn{1}{c}{$S_1 \to S_2$} &
\multicolumn{1}{c}{$S_2 \to S_1$} &
\multicolumn{2}{c}{$S_1 \to S_2$} &
\multicolumn{2}{c}{$S_2 \to S_1$} \\
\midrule
C15
 & $8b$    & (\ce{B4})   & $32e$   & (\ce{AB3})  & \ce{B3}  & B--H & A--H & 196 & (186) & 739 & (801) \\
 & $32e$   & (\ce{AB3})  & $96g$   & (\ce{A2B2}) & \ce{AB2} & B--H & A--H & 119 &  (90) & 288 & (290) \\
 & $96g$   & (\ce{A2B2}) & $96g$   & (\ce{A2B2}) & \ce{AB2} & A--H & A--H & 282 & (285) & 282 & (285) \\
 & $96g$   & (\ce{A2B2}) & $96g$   & (\ce{A2B2}) & \ce{A2B} & B--H & B--H & 138 & (123) & 138 & (123) \\
\midrule
C14
 & $4e$    & (\ce{B4})   & $4e$    & (\ce{B4})   & \ce{B3}  & B--H & B--H & 189 & (154) & 189 & (154) \\
 & $4e$    & (\ce{B4})   & $12k_1$ & (\ce{AB3})  & \ce{B3}  & B--H & A--H & 142 & (154) & 711 & (739) \\
 & $4f$    & (\ce{AB3})  & $4f$    & (\ce{AB3})  & \ce{B3}  & A--H & A--H & 695 & (630) & 695 & (630) \\
 & $4f$    & (\ce{AB3})  & $12k_2$ & (\ce{A2B2}) & \ce{AB2} & B--H & A--H &  105 & (111) & 343 & (361) \\
 & $12k_1$ & (\ce{AB3})  & $6h_1$  & (\ce{A2B2}) & \ce{AB2} & B--H & A--H &  106 &  (98) & 253 & (249) \\
 & $12k_1$ & (\ce{AB3})  & $24l$   & (\ce{A2B2}) & \ce{AB2} & B--H & A--H &  107 &  (97) & 269 & (263) \\
 & $6h_1$  & (\ce{A2B2}) & $6h_2$  & (\ce{A2B2}) & \ce{A2B} & B--H & B--H & 100 &  (87) & 157 & (166) \\
 & $6h_2$  & (\ce{A2B2}) & $12k_2$ & (\ce{A2B2}) & \ce{AB2} & A--H & A--H & 305 & (280) & 219 & (203) \\
 & $12k_2$ & (\ce{A2B2}) & $24l$   & (\ce{A2B2}) & \ce{A2B} & B--H & B--H & 111 & (111) & 155 & (124) \\
 & $24l$   & (\ce{A2B2}) & $24l$   & (\ce{A2B2}) & \ce{AB2} & A--H & A--H & 272 & (313) & 272 & (313) \\
 & $24l$   & (\ce{A2B2}) & $24l$   & (\ce{A2B2}) & \ce{A2B} & B--H & B--H & 122 &  (100) & 122 &  (100) \\
\bottomrule
\end{tabular}

\end{table*}

\color{black}

\section{Evolution of mean squared displacements}
\label{sec:MSD_all}

\begin{figure*}[htbp]
\centering
\includegraphics{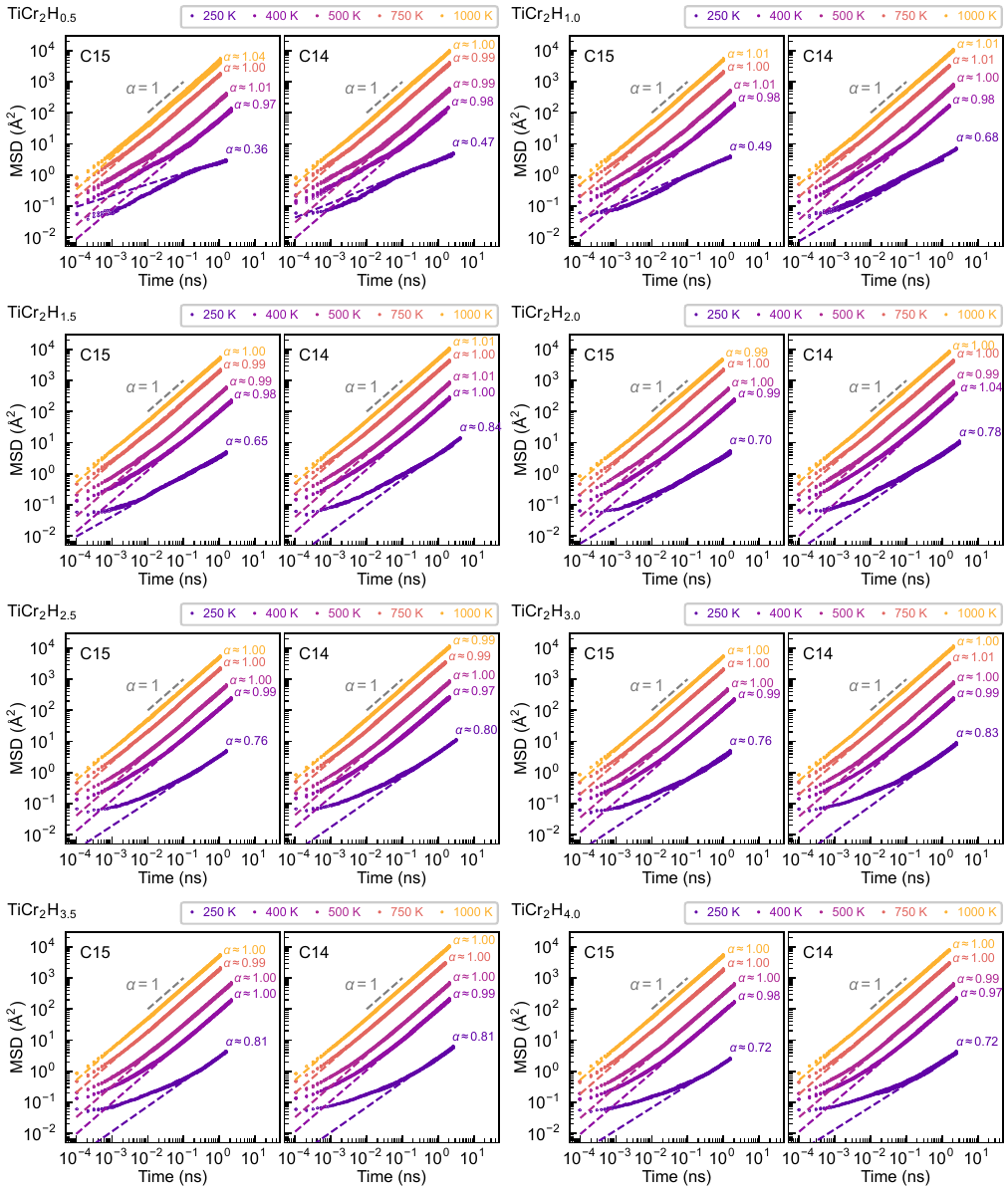}
\caption{MSDs of hydrogen atoms in \ce{TiCr2H_x} as a function of MD simulation time.}
\label{fig:MSD_all}
\end{figure*}

Figure~\ref{fig:MSD_all} illustrates the evolution of the MSD of hydrogen in \ce{TiCr2H_x} over MD time at five different temperatures: \qtylist{250;400;500;750;1000}{K}.

\section{Diffusion coefficients}

Tables~\ref{tab:diffusion:C15} and \ref{tab:diffusion:C14} present the diffusion coefficients of \ce{TiCr2H_x} obtained with the MD simulations.

\begin{table*}[!htbp]
\centering
\scriptsize
\caption{Diffusion coefficients (\qty{e-7}{cm^2/s}) of C15 \ce{TiCr2H_x} obtained with the MD simulations.}
\label{tab:diffusion:C15}
\begin{tabular}{ccS[table-format=1.1,table-auto-round]*7{S[table-format=3.3]}}
\toprule
 & $T$ (K) & $x$ & 
\multicolumn{1}{c}{MD1} &
\multicolumn{1}{c}{MD2} &
\multicolumn{1}{c}{MD3} &
\multicolumn{1}{c}{MD4} &
\multicolumn{1}{c}{MD5} &
\multicolumn{1}{c}{Mean} &
\multicolumn{1}{c}{SD} \\
\midrule
C15 & 400 & 0.5 & 10.133 & 11.256 & 10.828 & 10.612 & 10.818 & 10.729 & 0.364 \\
 &  & 1.0 & 15.779 & 15.118 & 14.691 & 15.142 & 14.508 & 15.047 & 0.440 \\
 &  & 1.5 & 19.651 & 18.644 & 19.541 & 18.708 & 18.333 & 18.975 & 0.523 \\
 &  & 2.0 & 19.931 & 20.669 & 21.436 & 20.972 & 21.208 & 20.843 & 0.522 \\
 &  & 2.5 & 20.335 & 19.910 & 20.154 & 19.876 & 20.920 & 20.239 & 0.380 \\
 &  & 3.0 & 17.759 & 17.562 & 18.469 & 17.588 & 18.425 & 17.961 & 0.403 \\
 &  & 3.5 & 16.132 & 16.327 & 15.682 & 15.371 & 15.470 & 15.797 & 0.373 \\
 &  & 4.0 & 13.514 & 14.194 & 13.462 & 13.482 & 13.404 & 13.611 & 0.293 \\
\cmidrule{2-10}
 & 500 & 0.5 & 43.836 & 47.473 & 45.115 & 46.148 & 46.425 & 45.799 & 1.236 \\
 &  & 1.0 & 54.713 & 54.713 & 56.394 & 57.251 & 53.628 & 55.340 & 1.302 \\
 &  & 1.5 & 65.037 & 64.159 & 63.873 & 65.474 & 66.249 & 64.958 & 0.867 \\
 &  & 2.0 & 67.204 & 67.216 & 68.401 & 67.204 & 67.216 & 67.448 & 0.476 \\
 &  & 2.5 & 66.011 & 67.061 & 64.694 & 68.573 & 65.918 & 66.451 & 1.299 \\
 &  & 3.0 & 60.746 & 61.002 & 60.497 & 57.662 & 60.344 & 60.050 & 1.215 \\
 &  & 3.5 & 54.468 & 56.515 & 57.206 & 54.342 & 56.848 & 55.876 & 1.221 \\
 &  & 4.0 & 51.039 & 52.630 & 53.022 & 51.430 & 51.375 & 51.899 & 0.779 \\
\cmidrule{2-10}
 & 750 & 0.5 & 294.454 & 300.831 & 305.091 & 299.971 & 288.792 & 297.828 & 5.646 \\
 &  & 1.0 & 330.314 & 338.418 & 332.833 & 338.007 & 320.336 & 331.982 & 6.585 \\
 &  & 1.5 & 350.125 & 367.585 & 363.297 & 348.385 & 352.306 & 356.340 & 7.655 \\
 &  & 2.0 & 371.810 & 368.607 & 377.147 & 367.596 & 363.333 & 369.699 & 4.606 \\
 &  & 2.5 & 367.617 & 354.307 & 363.309 & 356.141 & 360.240 & 360.323 & 4.813 \\
 &  & 3.0 & 345.693 & 351.547 & 346.552 & 347.483 & 338.582 & 345.972 & 4.204 \\
 &  & 3.5 & 322.819 & 331.094 & 319.180 & 318.423 & 337.587 & 325.821 & 7.404 \\
 &  & 4.0 & 315.755 & 299.955 & 308.442 & 317.930 & 320.202 & 312.457 & 7.392 \\
\cmidrule{2-10}
 & 1000 & 0.5 & 801.825 & 829.650 & 832.483 & 802.807 & 871.537 & 827.660 & 25.446 \\
 &  & 1.0 & 863.952 & 856.831 & 856.839 & 848.338 & 886.754 & 862.543 & 13.077 \\
 &  & 1.5 & 898.824 & 904.457 & 891.629 & 877.511 & 903.267 & 895.138 & 9.895 \\
 &  & 2.0 & 926.771 & 900.716 & 897.442 & 902.577 & 909.486 & 907.398 & 10.456 \\
 &  & 2.5 & 922.420 & 900.865 & 893.941 & 911.371 & 904.076 & 906.535 & 9.721 \\
 &  & 3.0 & 894.984 & 880.986 & 900.187 & 883.554 & 898.665 & 891.675 & 7.905 \\
 &  & 3.5 & 867.791 & 887.523 & 898.665 & 870.310 & 896.538 & 884.165 & 12.920 \\
 &  & 4.0 & 888.874 & 875.564 & 894.710 & 845.373 & 891.865 & 879.277 & 18.177 \\
\bottomrule
\end{tabular}
\end{table*}

\begin{table*}[!htbp]
\centering
\scriptsize
\caption{Diffusion coefficients (\qty{e-7}{cm^2/s}) of C14 \ce{TiCr2H_x} obtained with the MD simulations.}
\label{tab:diffusion:C14}
\begin{tabular}{ccS[table-format=1.1,table-auto-round]*7{S[table-format=3.3]}}
\toprule
 & $T$ (K) & $x$ & 
\multicolumn{1}{c}{MD1} &
\multicolumn{1}{c}{MD2} &
\multicolumn{1}{c}{MD3} &
\multicolumn{1}{c}{MD4} &
\multicolumn{1}{c}{MD5} &
\multicolumn{1}{c}{Mean} &
\multicolumn{1}{c}{SD} \\
\midrule
C14 & 400 & 0.5 & 14.344 & 13.564 & 13.411 & 13.154 & 12.154 & 13.326 & 0.708 \\
 &  & 1.0 & 18.460 & 17.615 & 17.913 & 17.917 & 17.592 & 17.899 & 0.313 \\
 &  & 1.5 & 22.131 & 22.610 & 23.087 & 23.812 & 22.236 & 22.775 & 0.617 \\
 &  & 2.0 & 23.910 & 23.389 & 25.452 & 24.302 & 22.825 & 23.976 & 0.890 \\
 &  & 2.5 & 22.056 & 24.177 & 21.610 & 22.654 & 22.233 & 22.546 & 0.882 \\
 &  & 3.0 & 19.256 & 19.484 & 20.187 & 19.501 & 18.795 & 19.445 & 0.450 \\
 &  & 3.5 & 15.925 & 16.742 & 16.766 & 17.140 & 17.523 & 16.819 & 0.530 \\
 &  & 4.0 & 15.068 & 14.066 & 14.423 & 14.875 & 14.847 & 14.656 & 0.362 \\
\cmidrule{2-10}
 & 500 & 0.5 & 46.090 & 51.443 & 47.042 & 45.522 & 46.270 & 47.273 & 2.141 \\
 &  & 1.0 & 67.638 & 62.584 & 64.206 & 61.558 & 62.591 & 63.715 & 2.137 \\
 &  & 1.5 & 72.044 & 74.203 & 72.824 & 71.802 & 70.507 & 72.276 & 1.218 \\
 &  & 2.0 & 80.570 & 79.830 & 77.043 & 75.387 & 71.707 & 76.908 & 3.204 \\
 &  & 2.5 & 72.236 & 68.354 & 70.882 & 72.121 & 71.528 & 71.024 & 1.419 \\
 &  & 3.0 & 61.677 & 64.010 & 63.500 & 62.451 & 61.301 & 62.588 & 1.035 \\
 &  & 3.5 & 57.026 & 57.653 & 56.352 & 55.087 & 55.507 & 56.325 & 0.944 \\
 &  & 4.0 & 50.857 & 52.983 & 53.515 & 50.094 & 50.435 & 51.577 & 1.396 \\
\cmidrule{2-10}
 & 750 & 0.5 & 315.832 & 326.092 & 329.621 & 338.672 & 319.363 & 325.916 & 8.014 \\
 &  & 1.0 & 366.231 & 349.263 & 364.663 & 352.065 & 353.294 & 357.103 & 6.955 \\
 &  & 1.5 & 376.589 & 377.106 & 375.775 & 380.565 & 381.136 & 378.234 & 2.185 \\
 &  & 2.0 & 381.543 & 371.881 & 372.906 & 382.447 & 369.274 & 375.610 & 5.354 \\
 &  & 2.5 & 379.671 & 364.795 & 366.926 & 380.577 & 373.034 & 373.001 & 6.421 \\
 &  & 3.0 & 351.194 & 351.553 & 353.284 & 353.786 & 344.733 & 350.910 & 3.242 \\
 &  & 3.5 & 326.387 & 328.774 & 342.707 & 335.671 & 336.342 & 333.976 & 5.820 \\
 &  & 4.0 & 309.857 & 320.603 & 315.014 & 323.485 & 325.319 & 318.856 & 5.692 \\
\cmidrule{2-10}
 & 1000 & 0.5 & 833.150 & 827.726 & 834.807 & 830.568 & 829.881 & 831.226 & 2.491 \\
 &  & 1.0 & 887.020 & 896.067 & 900.972 & 884.110 & 855.862 & 884.806 & 15.692 \\
 &  & 1.5 & 912.921 & 921.323 & 894.645 & 888.604 & 896.125 & 902.724 & 12.313 \\
 &  & 2.0 & 935.324 & 937.033 & 915.019 & 937.695 & 908.690 & 926.752 & 12.352 \\
 &  & 2.5 & 918.788 & 894.380 & 895.554 & 931.341 & 916.876 & 911.388 & 14.304 \\
 &  & 3.0 & 887.772 & 914.446 & 906.416 & 920.082 & 900.757 & 905.895 & 11.221 \\
 &  & 3.5 & 865.096 & 895.785 & 900.924 & 885.466 & 895.355 & 888.525 & 12.739 \\
 &  & 4.0 & 898.837 & 868.894 & 885.213 & 885.991 & 894.711 & 886.729 & 10.308 \\
\bottomrule
\end{tabular}
\end{table*}

\newpage
\bibliographystyle{elsarticle-num-mod} 
\bibliography{Supplementary.bbl}